\RequirePackage{fix-cm}

\documentclass[twocolumn,epjc3]{svjour3} 
\smartqed  
\RequirePackage{graphicx}

\RequirePackage[colorlinks,citecolor=blue,urlcolor=blue,linkcolor=blue]{hyperref}
\journalname{}
\begin{document}

\title{Neutral bremsstrahlung and excimer electroluminescence in noble gases and its relevance to two-phase dark matter detectors
	}


\author{E. Borisova\thanksref{addr1,addr2}
        \and 
        A. Buzulutskov\thanksref{addr1,addr2,e1}
        }

\thankstext{e1}{A.F.Buzulutskov@inp.nsk.su (corresponding author)}


\institute{Budker Institute of Nuclear Physics SB RAS, Lavrentiev avenue 11, 630090 Novosibirsk, Russia \label{addr1} 
\and Novosibirsk State University, Pirogov street 2, 630090 Novosibirsk, Russia \label{addr2}
}

\date{}

\maketitle

\begin{abstract}
Proportional electroluminescence (EL) is the physical effect used in two-phase detectors for dark matter searches, to optically record (in the gas phase) the ionization signal produced by particle scattering in the liquid phase. In our previous work the presence of a new EL mechanism, namely that of neutral bremsstrahlung (NBrS), was demonstrated in two-phase argon detectors both theoretically and experimentally, in addition to the ordinary EL mechanism due to excimer emission. In this work the similar theoretical approach is applied to all noble gases, i.e. overall to helium, neon, argon, krypton and xenon, to calculate the EL yields and spectra both for NBrS and excimer EL.  The relevance of the results obtained to the development of two-phase dark matter detectors is discussed.
\end{abstract}

\section{Introduction}

In two-phase detectors for dark matter searches and low-energy neutrino experiments, the scattered particle produces two types of signals~\cite{Chepel13}: that of primary scintillation, produced in the liquid and recorded promptly (``S1''), and that of primary ionization, produced in the liquid  and recorded with a delay in the gas phase (``S2''). To record the S2 signal, proportional electroluminescence (EL) is used, produced by drifting electrons under high enough electric fields. 

According to modern concepts~\cite{Buzulutskov20}, there are three mechanisms responsible for proportional EL in noble gases: that of excimer (e.g. Ar$^*_2$) emission in the vacuum ultraviolet (VUV), that of emission due to atomic transitions in the near infrared (NIR), and that of neutral bremsstrahlung (NBrS) emission in the UV, visible and NIR range. In the following these three mechanisms are referred to as excimer (or else ordinary) EL, atomic EL and NBrS EL, respectively. 

The presence of NBrS EL in two-phase Ar detectors has for the first time been demonstrated in our previous work~\cite{Buzulutskov18}, both theoretically and experimentally. 
In particular it was shown that the NBrS effect can explain two intriguing observations in EL radiation of gaseous Ar: that of the substantial contribution of the non-VUV spectral component (in the UV, visible and NIR range), and that of the photon emission at lower electric fields, below the Ar excitation threshold. The merit of that work was that it transformed the idea of NBrS EL from a hypothesis~\cite{Butikov70} into a quantitative theory~\cite{Buzulutskov18}. The success of the NBrS theory developed there was that it correctly predicted the absolute value of the EL yield below the Ar excitation threshold. 

In this work, the similar theoretical approach is applied to all noble gases both for NBrS and excimer EL, to calculate the EL yields and spectra: to He, Ne, Ar, Kr and Xe. The relevance of the results obtained to the development of two-phase dark matter detectors is also discussed.

\section{Overview of electroluminescence (EL) mechanisms in noble gases}

\begin{figure}[htb]
	\centering
	\includegraphics[width=0.99\columnwidth,keepaspectratio]{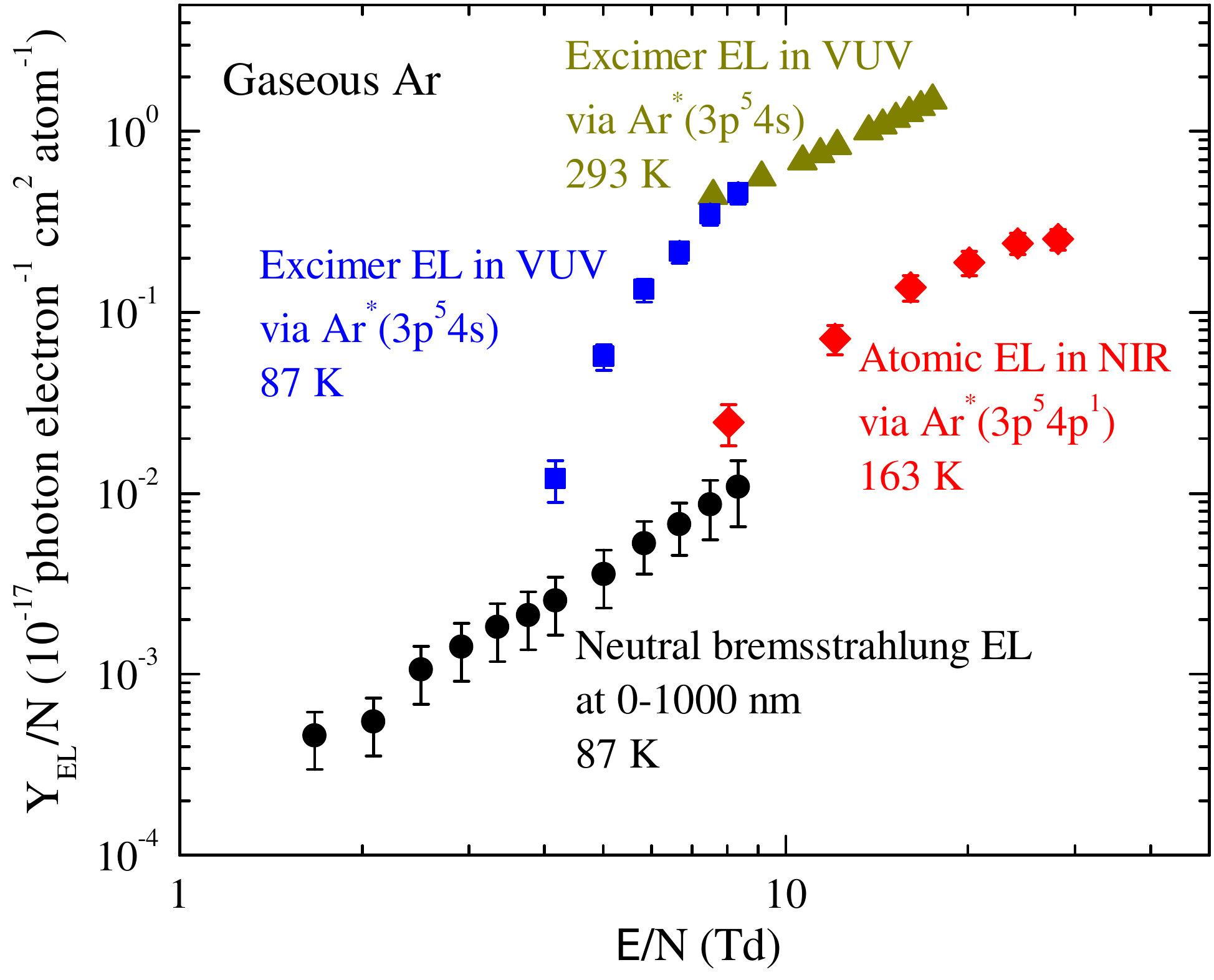}
	\caption{Summary of experimental data on reduced electroluminescence (EL) yield in gaseous Ar for all known EL mechanisms: for neutral bremsstrahlung (NBrS) EL at wavelengths of 0-1000 nm obtained at 87~K in~\cite{Buzulutskov18,Bondar20}; for excimer EL in the VUV going via Ar$^{\ast}(3p^54s)$ excited states, obtained at 87~K in~\cite{Bondar20} and at 293~K in \cite{Monteiro08} (the data points of the latter below 7 Td are not shown as they include both excimer and NBrS EL); for EL in the NIR, going via Ar$^{\ast}(3p^54p)$ excited sates, obtained at 163~K in~\cite{Buzulutskov11}. For NBrS EL at $\mathcal{E}/N>$~4~Td, the experimental EL yield exceeds that predicted by the theory for standard NBrS EL; therefore the data points  at these fields might be somewhat incorrect, as they were obtained assuming the standard NBrS EL emission spectrum which might not be the case. }
	\label{Int01}
\end{figure}

\begin{figure}[htb]
	\centering
	\includegraphics[width=0.99\columnwidth,keepaspectratio]{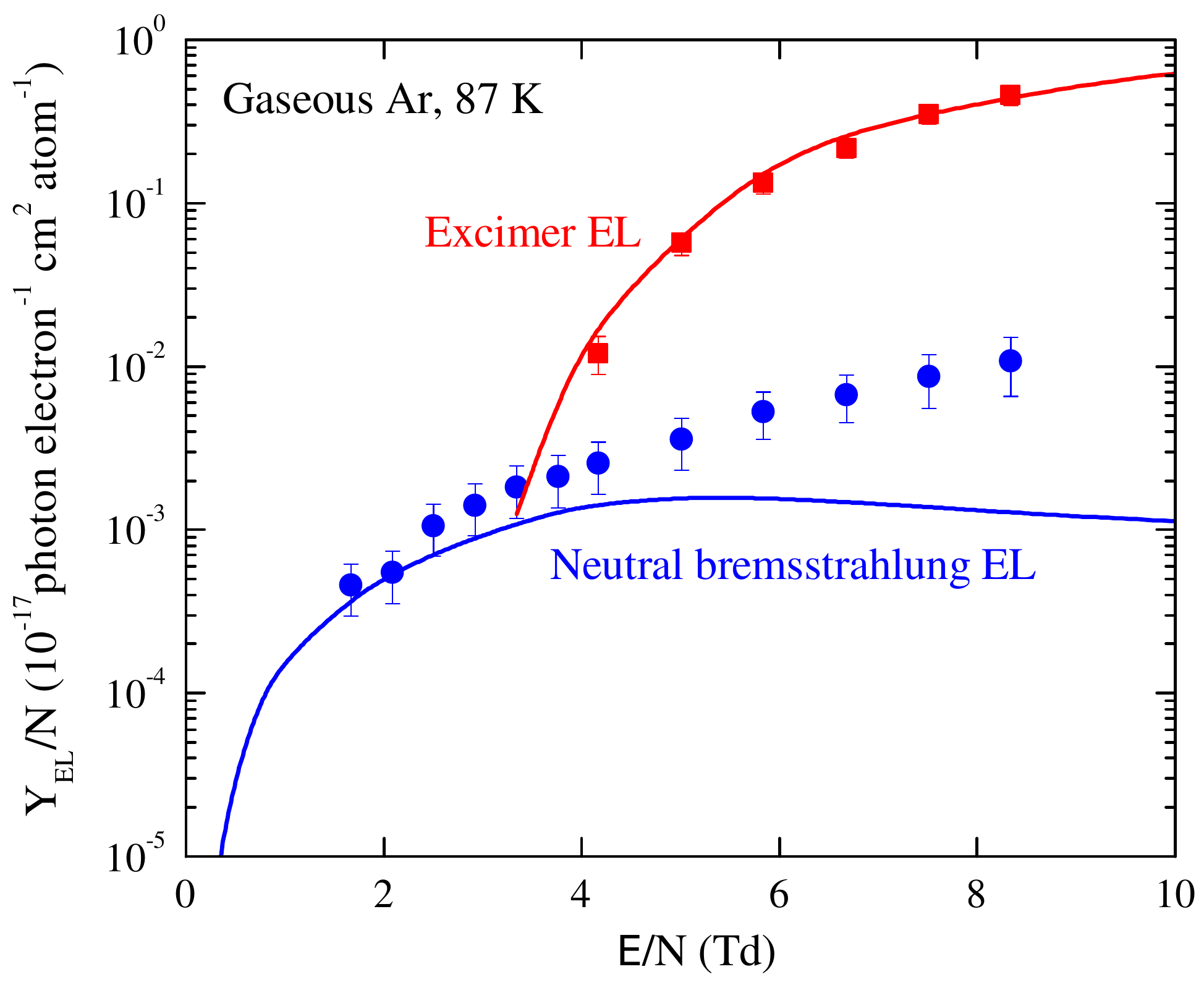}
	\caption{Experimental data (points) on reduced EL yield in gaseous Ar at lower reduced electric fields, below 9 Td, for NBrS EL \cite{Buzulutskov18,Bondar20} and excimer EL \cite{Bondar20}, in comparison with the theory (curves) presented in \cite{Buzulutskov18} and \cite{Oliveira11} respectively. For NBrS EL at $\mathcal{E}/N>$~4~Td, the experimental EL yield exceeds that predicted by the theory for standard NBrS EL; therefore the data points  at these fields might be somewhat incorrect, as they were obtained assuming the standard NBrS EL emission spectrum which might not be the case.}
	\label{Fig00Ar}
\end{figure}

The three EL mechanisms in noble gases are illustrated in Figs.~\ref{Int01} and \ref{Fig00Ar}, on the example of Ar, showing the dependence of the reduced EL yield on the reduced electric field. The first figure shows the experimentally measured yields, while the second one compares the experimental yields to those  calculated theoretically. 

Here the reduced EL yield, $Y_{EL}/N$, is defined as the number of photons produced per unit drift path ($dN_{ph}/dx$) and per drifting electron, normalized to the atomic density ($N$): 
\begin{eqnarray}
	\label{Eq-NBrS-el-yield} 
	\frac{Y_{EL}}{N} = \frac{dN_{ph}}{dx \ N \ N_e \ dV} 	\; , 
\end{eqnarray}
where $N_e$ is the density of drifting electrons, $dV$ is the volume and $N_e dV$ is thus the number of drifting electrons. The reduced electric field is defined as $\mathcal{E}/N$, where $\mathcal{E}$ is the electric field. It is expressed in Td units:  1~Td~=~$10^{-17}$~V~cm$^2$, corresponding to the electric field of 0.87~kV/cm in gaseous Ar in the two-phase mode at 87.3~K and 1.00 atm.

Excimer (ordinary) EL is due to emission of noble gas excimers, Ar$^{*}_{2}(^{1,3}\Sigma^{+}_{u})$, produced in three-body atomic collisions of excited atoms, Ar$^*(3p^54s)$, which in turn are produced by drifting electrons in electron-atom collisions (see reviews~\cite{Chepel13,Buzulutskov20,Buzulutskov17}):
\begin{eqnarray}
	\label{Rea-ord-el}
	e^- + \mathrm{Ar} \rightarrow e^- + \mathrm{Ar}^{\ast}(3p^54s) \; , \nonumber \\
	\mathrm{Ar}^{\ast}(3p^54s) + 2\mathrm{Ar} \rightarrow \mathrm{Ar}^{\ast}_2(^{1,3}\Sigma^{+}_{u}) + \mathrm{Ar} \; , \nonumber \\
	\mathrm{Ar}^{\ast}_2(^{1,3}\Sigma^{+}_{u}) \rightarrow 2\mathrm{Ar} + h\nu \; .
\end{eqnarray}

\begin{figure}[htb]
	\centering
	\includegraphics[width=0.99\columnwidth,keepaspectratio]{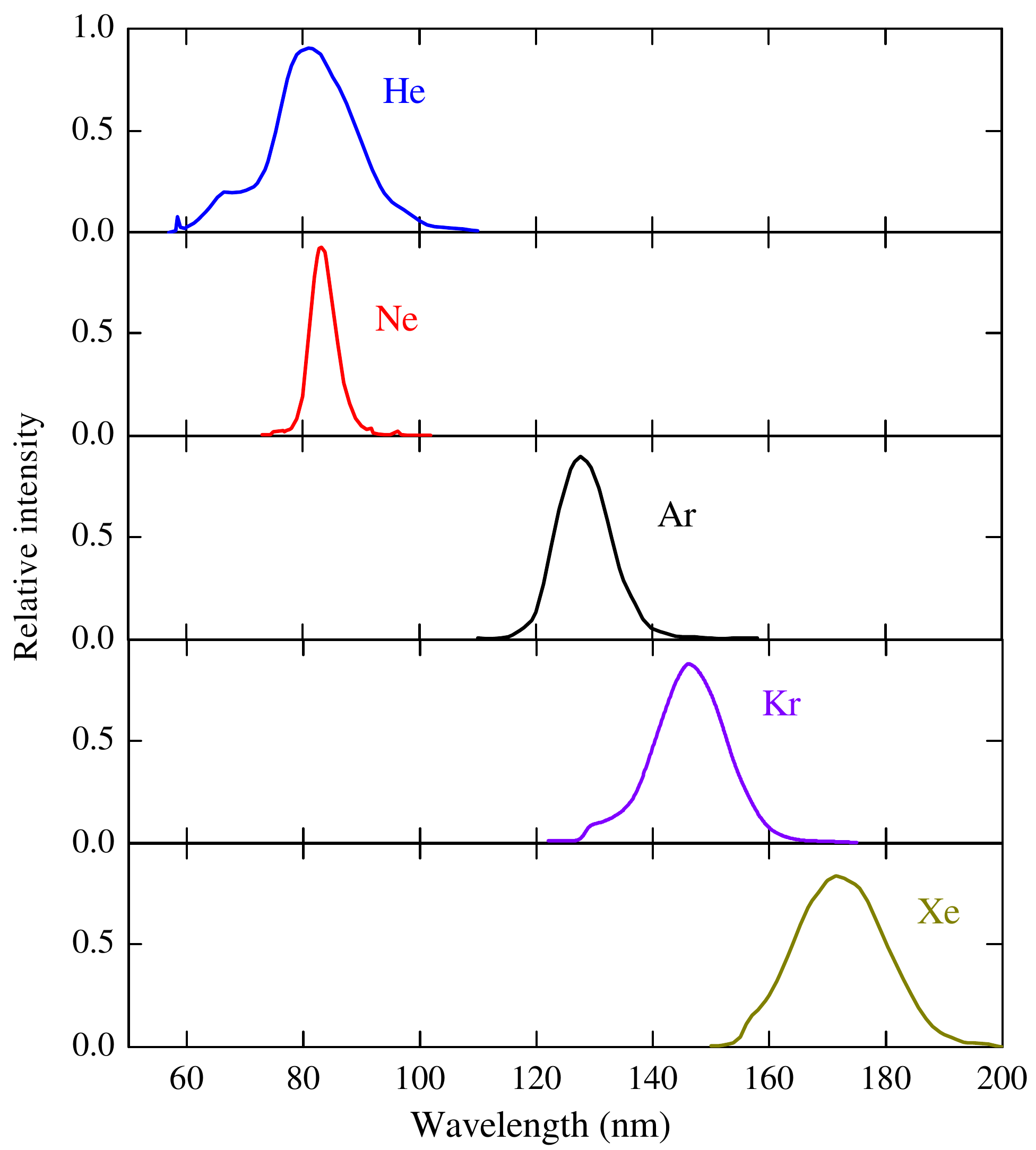}
	\caption{Spectra of the second VUV continuum in gaseous He, Ne, Ar, Kr and Xe due to excimer emission. In He it was induced by condensed discharge at a pressure of 40 Torr~\cite{Aprile06,Huffman65}, while in Ne, Ar, Kr and Xe it was induced by electron beam at a pressure of 120 kPa~\cite{Morozov08}.}
	\label{Int00}
\end{figure}

Excimer EL in noble gases has a threshold in the electric field (of about 4 Td for Ar), defined by the lowest atomic excitation levels Ar$^*(3p^54s)$.  In addition it is characterized by emission continuum in the VUV (so-called ``second continuum''~\cite{Schwentner85}): Fig.~\ref{Int00} shows this for all noble gases. From this figure one can see that excimer EL can be recorded directly only in Xe, using photomultiplier tubes (PMTs) or silicon photomultipliers (SiPMs) with quartz windows, while in other noble gases (He, Ne, Ar and Kr) it can be recorded indirectly (since quartz is not transparent below 160 nm) using a wavelength shifter (WLS), typically tetraphenyl-butadiene (TPB)~\cite{Benson18}.   

\begin{figure}[htb]
	\centering
	\includegraphics[width=0.99\columnwidth,keepaspectratio]{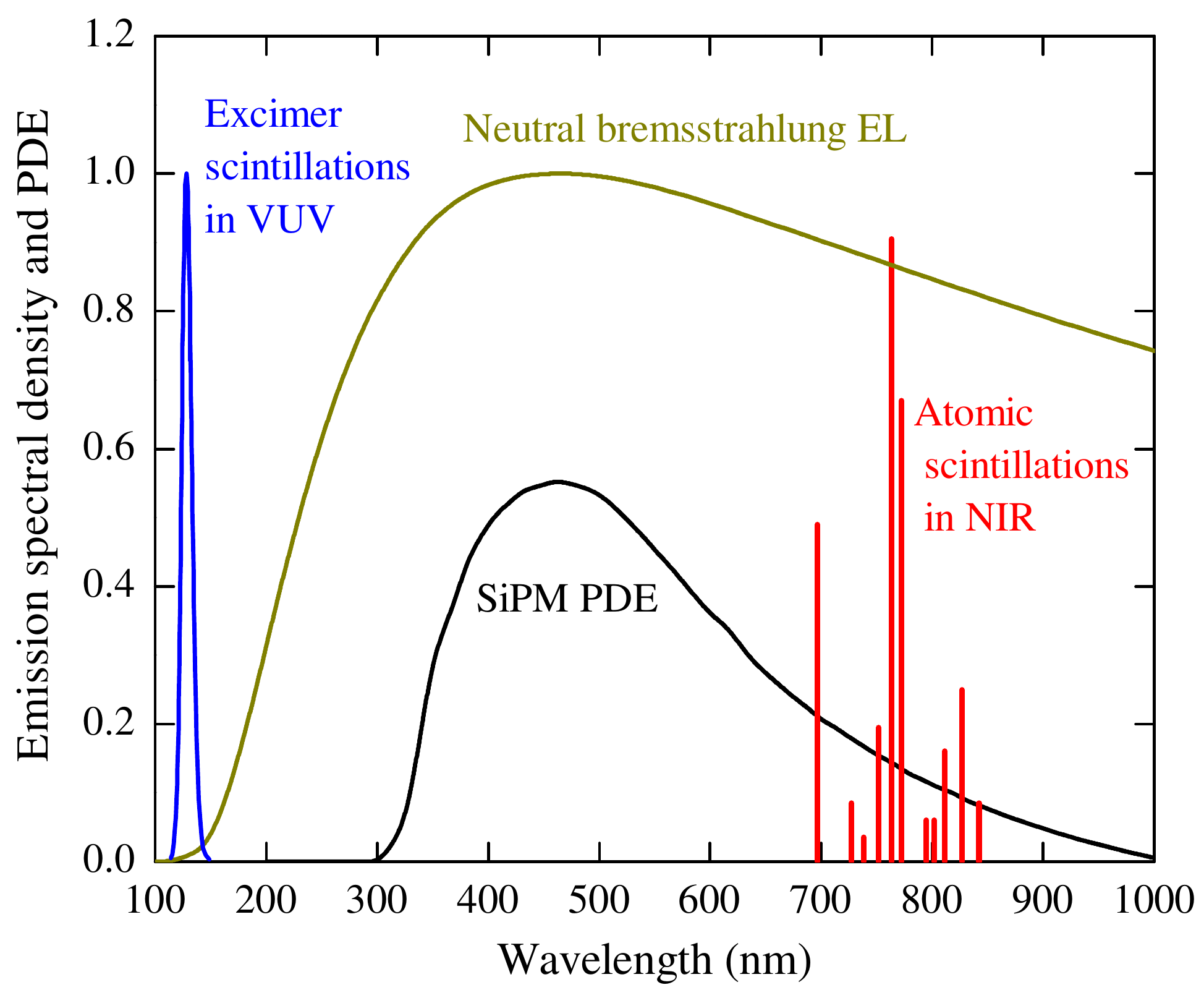}
	\caption{Photon emission spectra in gaseous Ar due to excimer scintillations in the VUV measured in~\cite{Morozov08}, NBrS EL at 8.3~Td theoretically calculated in~\cite{Buzulutskov18} and atomic scintillations in the NIR measured in \cite{Lindblom88,Fraga00}. Also shown is the photon detection efficiency (PDE) of the typical SiPM~\cite{Aalseth21}.}
	\label{FigArSpectra}
\end{figure}

Atomic EL in the NIR is due to atomic transitions between the excited states \cite{Buzulutskov11,Oliveira13}:
\begin{eqnarray}
\mathrm{Ar}^*(3p^54p) \rightarrow \mathrm{Ar}^*(3p^54s)+h\nu \; .
\end{eqnarray}
It has a line emission spectrum \cite{Buzulutskov20,Lindblom88} (see Fig.~\ref{FigArSpectra}) and a $\sim$1 Td higher electric field threshold compared to excimer EL, i.e. at about 5~Td~\cite{Oliveira13}. 

Neutral bremsstrahlung (NBrS) EL is due to bremsstrahlung of drifting electrons scattered (elastically or inelastically) on neutral atoms:
\begin{eqnarray}
\label{Rea-NBrS-el}
e^- + \mathrm{Ar} \rightarrow e^- + \mathrm{Ar} + h\nu \; , \\
\label{Rea-NBrS-exc}
e^- + \mathrm{Ar} \rightarrow e^- + \mathrm{Ar}^{\ast} + h\nu \;.
\end{eqnarray}
In the most elaborated way, NBrS EL was introduced in \cite{Buzulutskov18} applying both theoretical and experimental approach to explain two observations: the photon emission  below the Ar excitation threshold and the contribution of the non-VUV spectral component in proportional EL (extending from the UV to NIR). It was  further experimentally studied in \cite{Bondar20,Tanaka20,Kimura20,Takeda20,Takeda20a,Aoyama21}.   

The theory predicts that the contribution of NBrS EL due to elastic collisions (reaction \ref{Rea-NBrS-el}) is considerably larger than that of inelastic \cite{Buzulutskov18}, since the elastic cross section for electron-atom collision is considerably  larger than that of inelastic. This should be true for all noble gases. Accordingly, in this work we consider NBrS EL due to elastic collisions only. 

NBrS EL has a continuous emission spectrum, extending from the UV to the visible and NIR range: see Fig.~\ref{FigArSpectra}. From Figs.~\ref{Int01} and \ref{Fig00Ar} one can see that NBrS EL, albeit being significantly weaker than excimer EL above the Ar excitation threshold, has no threshold in electric field, in contrast to excimer EL, and thus dominates below the threshold. 

At lower electric fields, below 4 Td corresponding to Ar excitation threshold, the NBrS theory developed in \cite{Buzulutskov18} correctly predicts the absolute value of the EL yield. This is seen when comparing the experimental and theoretical EL yields in Fig.~\ref{Fig00Ar} and when comparing the experimental  and theoretical photon emission spectra~\cite{Buzulutskov18,Tanaka20,Aoyama21}. 

On the other hand, at higher fields, above 4 Td, the experimental EL yield quickly diverges from that of the theory \cite{Buzulutskov18}, exceeding that in the UV spectral range (below 400 nm) \cite{Tanaka20}. In \cite{Buzulutskov18} this excess of experimental data over theoretical prediction was proposed to be explained by the contribution of electron scattering on sub-excitation Feshbach resonances \cite{Schulz73} (going via intermediate negative ion states Ar$^-(3p^54s^2)$), which might be accompanied by enhanced photon emission \cite{Buzulutskov18,Dyachkov74}:
\begin{eqnarray}
\label{NBrS-Res}
e^- + \mathrm{Ar} \rightarrow \mathrm{Ar}^-(3p^54s^2) \rightarrow e^- + \mathrm{Ar} + h\nu \; .
\end{eqnarray}
There are many negative ion resonances  above the Ar excitation thresold other than those of Feshbach \cite{Schulz73}, which may also give rise to photon emission similar to Eq.~\ref{NBrS-Res}. 
The energies of Feshbach resonances are close to that of the lowest excitation level (see Table~\ref{tbl:table1}, item 4 and 5), which should result in a field dependence similar to that of excimer EL, namely in the linear growth of the intensity with the electric field started at almost the same thresold as that of excimer EL (i.e. at 4 Td). 
Since the photon emission spectrum of this mechanism can differ from that of ``standard'' NBrS EL, the experimental data points in Figs.~\ref{Int01} and \ref{Fig00Ar} at $\mathcal{E}/N>$~4~Td might be somewhat incorrect, as they were obtained assuming the standard NBrS EL emission spectrum which might not be the case. Note that sub-excitation Feshbach resonances exist in all noble gases: see Table~\ref{tbl:table1}~(item 4) for their description and energies. Thus this mechanism (neutral bremsstrahlung on resonances) can be applied to all noble gases, if any.

An alternative explanation of the observed excess in EL yield was considered in \cite{Aoyama21}, using experimental data on the EL spectra in the visible range in pure Ar and its mixtures with N$_2$ obtained at room temperature: the idea that emission of N$_2$ impurity might be responsible for this excess was considered. In particular, it was observed that at a N$_2$ content of 100 ppm there was a significant contribution of characteristic emission peaks of N$_2^{\ast}(C)$ excited states (in the range of 300-400 nm \cite{Buzulutskov17}), which enhanced the overall light intensity in the range of 300-600 nm by about a factor of 2 compared to pure Ar (at 8 Td). However, for lower N$_2$ content, of 10 ppm, the spectrum almost did not differ from that of pure Ar. Since the claimed N$_2$ impurity content in the experimental data of Fig.~\ref{Fig00Ar} was below 1 ppm \cite{Buzulutskov18,Bondar20}, it is difficult to explain the discussed excess by the results of \cite{Aoyama21}, even if to involve the hypothetical mechanism of the enhancement of N$_2^{\ast}(C)$ production at low temperatures proposed elsewhere \cite{Buzulutskov17}. Obviously, further research is needed to clarify this issue.

NBrS EL is universal in nature: it must be present in all noble  gases. In particular, just recently the observation of NBrS EL in Xe and Kr below the atomic excitation threshold has been reported~\cite{Monteiro21}. In the following sections we apply the theoretical method developed in~\cite{Buzulutskov18} for Ar to all other noble gases, i.e. overall to He, Ne, Ar, Kr and Xe, both for NBrS and excimer EL.

\section{Theoretical formulas}

The differential cross section for NBrS photon emission is expressed via electron-atom elastic cross section ($\sigma _{el}(E)$)~\cite{Buzulutskov18,Park00,Firsov61,Kasyanov65,Dalgarno66,Biberman67}:
\begin{eqnarray}
	\label{Eq-sigma-el} 
	\frac{d\sigma}{d\nu} = \frac{8}{3} \frac{r_e}{c} \frac{1}{h\nu} \left(\frac{E - h\nu}{E} \right)^{1/2} \times \hspace{40pt} \nonumber \\ \times \ [(E-h\nu) \ \sigma _{el}(E) \ + \ E \ \sigma _{el}(E - h\nu) ]  \; ,
\end{eqnarray}
where $r_e=e^2/m_e c^2$ is the classical electron radius, $c=\nu \lambda$ is the speed of light, $E$ is the initial electron energy and $h\nu$ is the photon energy. 

The reduced EL yield ($Y_{EL}/N$) for NBrS EL can be described by the following equation \cite{Buzulutskov18}: 
\begin{eqnarray}
	\label{Eq-NBrS-el-yield} 
	\left( \frac{Y_{EL}}{N}\right)_{NBrS} =  \int\limits_{\lambda_1}^{\lambda_2}  \int\limits_{h\nu}^{\infty}\frac{\upsilon_e}{\upsilon_d} 
	\frac{d\sigma}{d\nu} \frac{d\nu}{d\lambda} f(E) \ dE \ d\lambda 
	\; , 
\end{eqnarray}
where $\upsilon_e=\sqrt{2E/m_e}$ is the electron velocity of chaotic motion,  $\upsilon_d$ is the electron drift velocity, $\lambda_1-\lambda_2$ is the sensitivity region of the photon detector,  
$d\nu/d\lambda=-c/\lambda^2$, $f(E)$ is the electron energy distribution function normalized as
\begin{eqnarray}
	\label{Eq-norm-f} 
	\int\limits_{0}^{\infty} f(E) \ dE = 1 \; .
\end{eqnarray}

Consequently, the spectrum of the reduced EL yield is 
\begin{eqnarray}
	\label{Eq-NBrS-el-yield-spectrum} 
	\frac{d (Y_{EL}/N)_{NBrS}}{d\lambda} = 
	\int\limits_{h\nu}^{\infty}\frac{\upsilon_e}{\upsilon_d} 
	\frac{d\sigma}{d\nu} \frac{d\nu}{d\lambda} f(E) \ dE \  
	\; . 
\end{eqnarray}

The reduced EL yield ($Y_{EL}/N$) for excimer EL can be described by the following equation \cite{Buzulutskov18}: 
\begin{eqnarray}
	\label{Eq-ord-el-yield}
	\left( \frac{Y_{EL}}{N}\right)_{excimer} =  \int\limits_{E_{exc}}^{\infty}\frac{\upsilon_e}{\upsilon_d} \sigma_{exc}(E) f(E) \ dE 
	\; , 
\end{eqnarray}
where $\sigma_{exc}(E)$ is the inelastic cross-section to produce an excited state in electron-atom collisions. Similarly to \cite{Oliveira11}, it is assumed that one excited state (e.g. Ar$^{\ast}$) produces one excimer state (e.g. Ar$^{\ast}_2$) and that one excimer produces one VUV photon.

\section{Cross sections and electron energy distribution functions}

The electron energy distribution functions were calculated by solving the Boltzmann equation using BOLSIG+ free software~\cite{Bolsig1,Bolsig2}, at a temperature corresponding to the boiling point at 1.0 atm for a given noble gas (see Table~\ref{tbl:table1},~item 1). The electron scattering cross sections from Biagi database ~\cite{DBBiagi}, transcribed from code Magboltz (version 8.9 for Ne or 8.97 for other gases)~\cite{Biagi99}, were used  as input data.

\begin{figure*}[!hp]
	\begin{minipage}[h]{0.49\linewidth}
		\center{\includegraphics[width=1\linewidth]{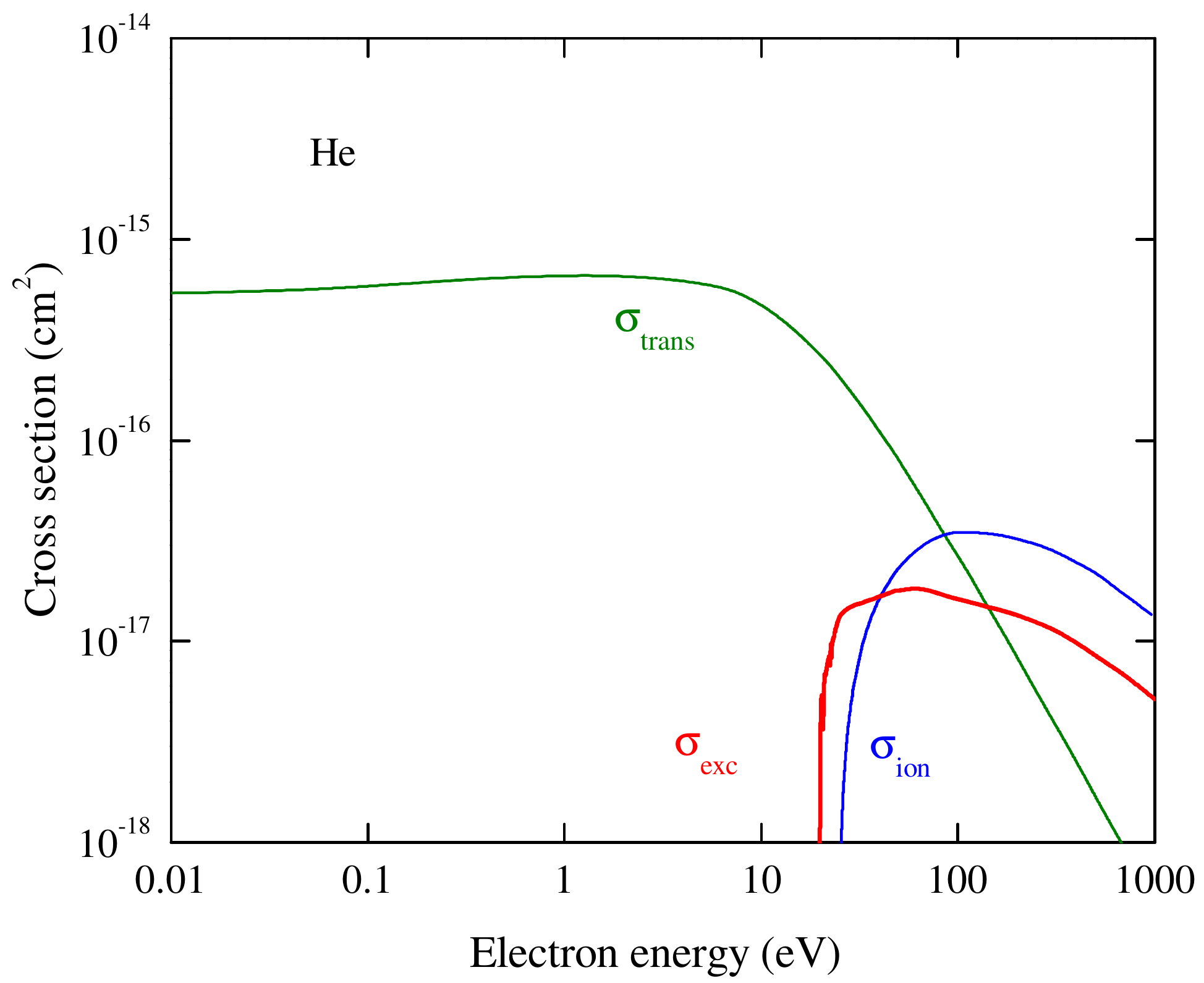}} \\
	\end{minipage}
	\hfill
	\begin{minipage}[h]{0.49\linewidth}
		\center{\includegraphics[width=1\linewidth]{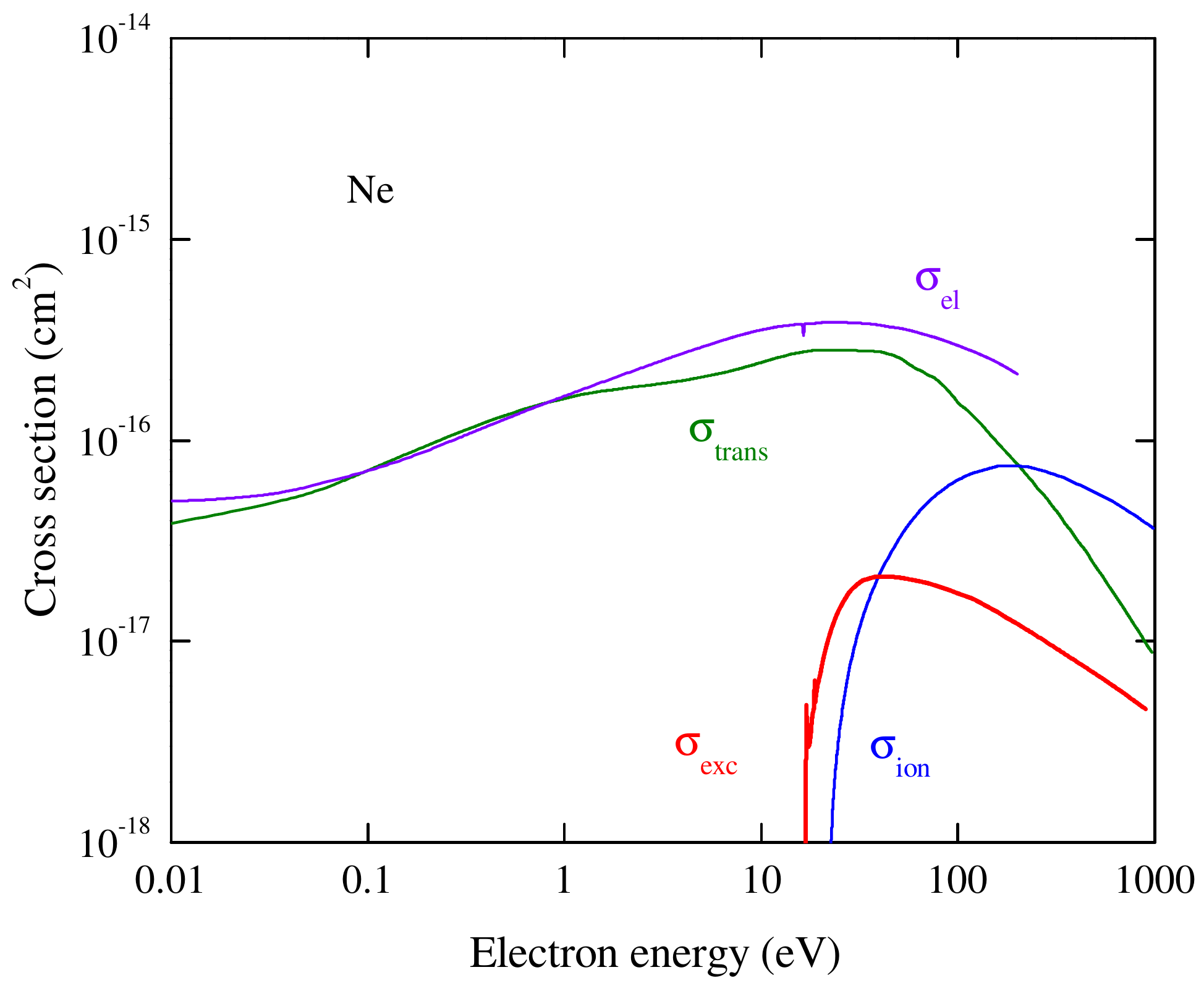}}
	\end{minipage}
	\vfill
	\begin{minipage}[h]{0.49\linewidth}
		\center{\includegraphics[width=1\linewidth]{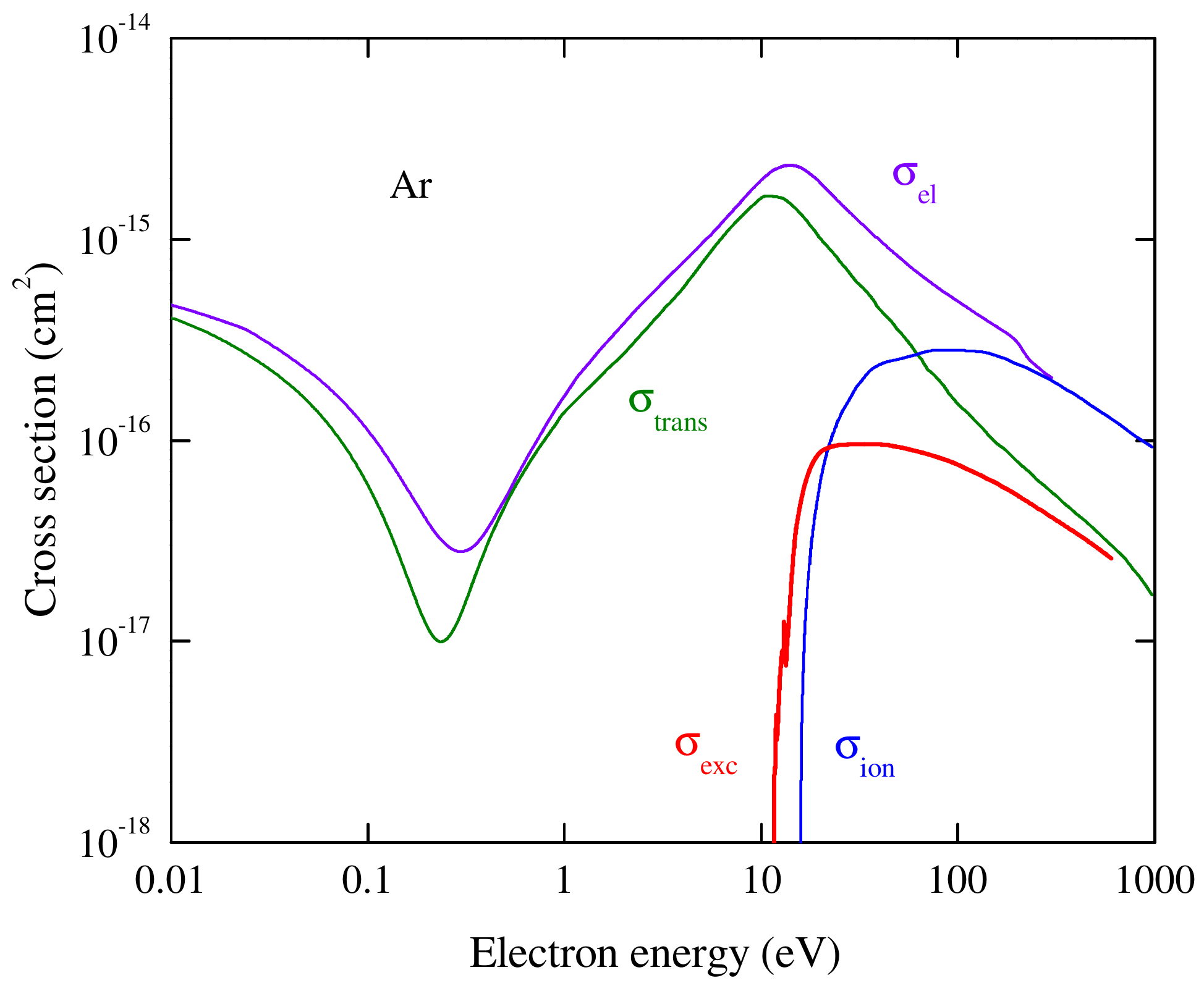}}\\
	\end{minipage}
	\hfill
	\begin{minipage}[h]{0.49\linewidth}
		\center{\includegraphics[width=1\linewidth]{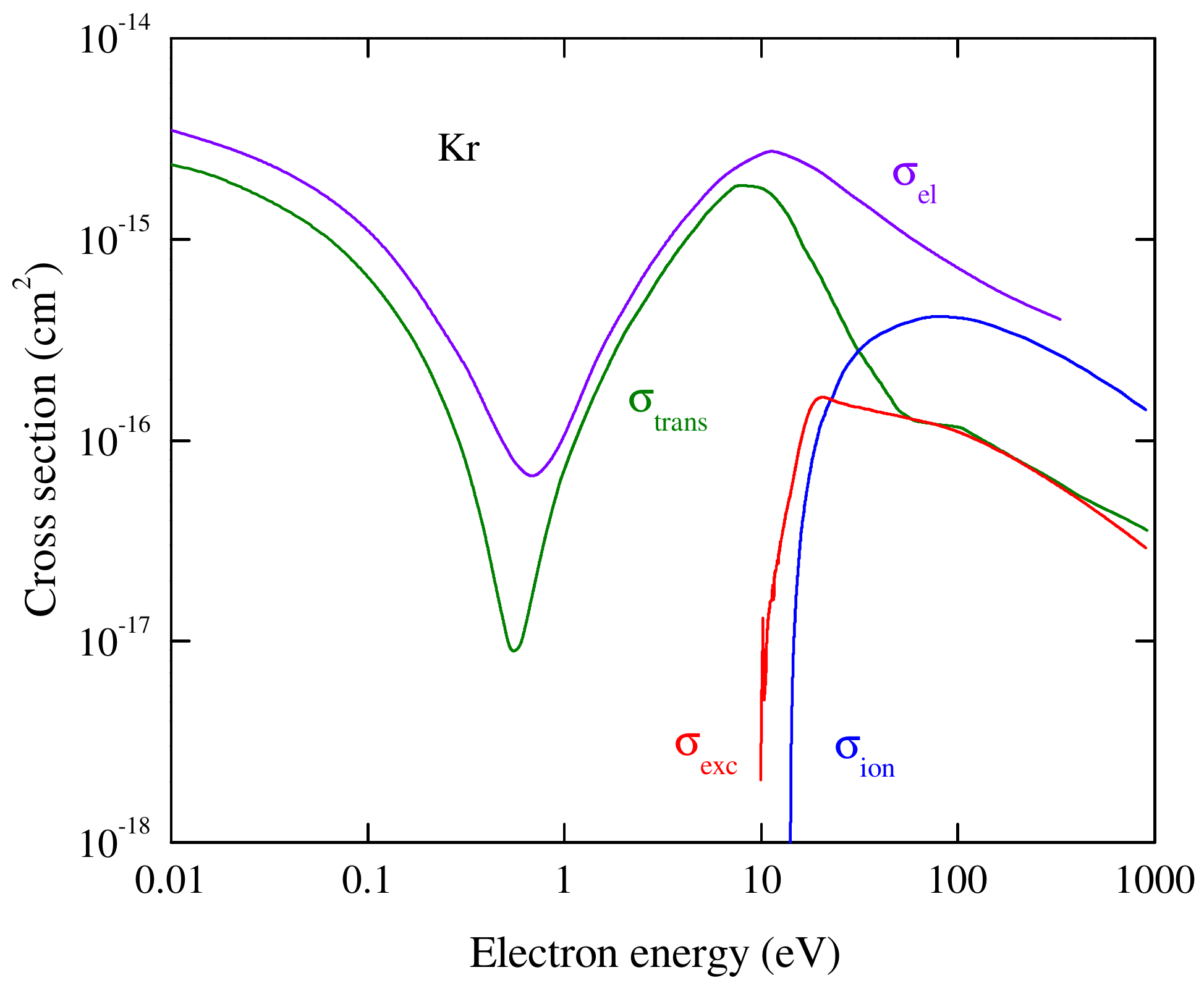}}\\
	\end{minipage}
	\vfill
	\begin{minipage}[h]{0.49\linewidth}
		\center{\includegraphics[width=1\linewidth]{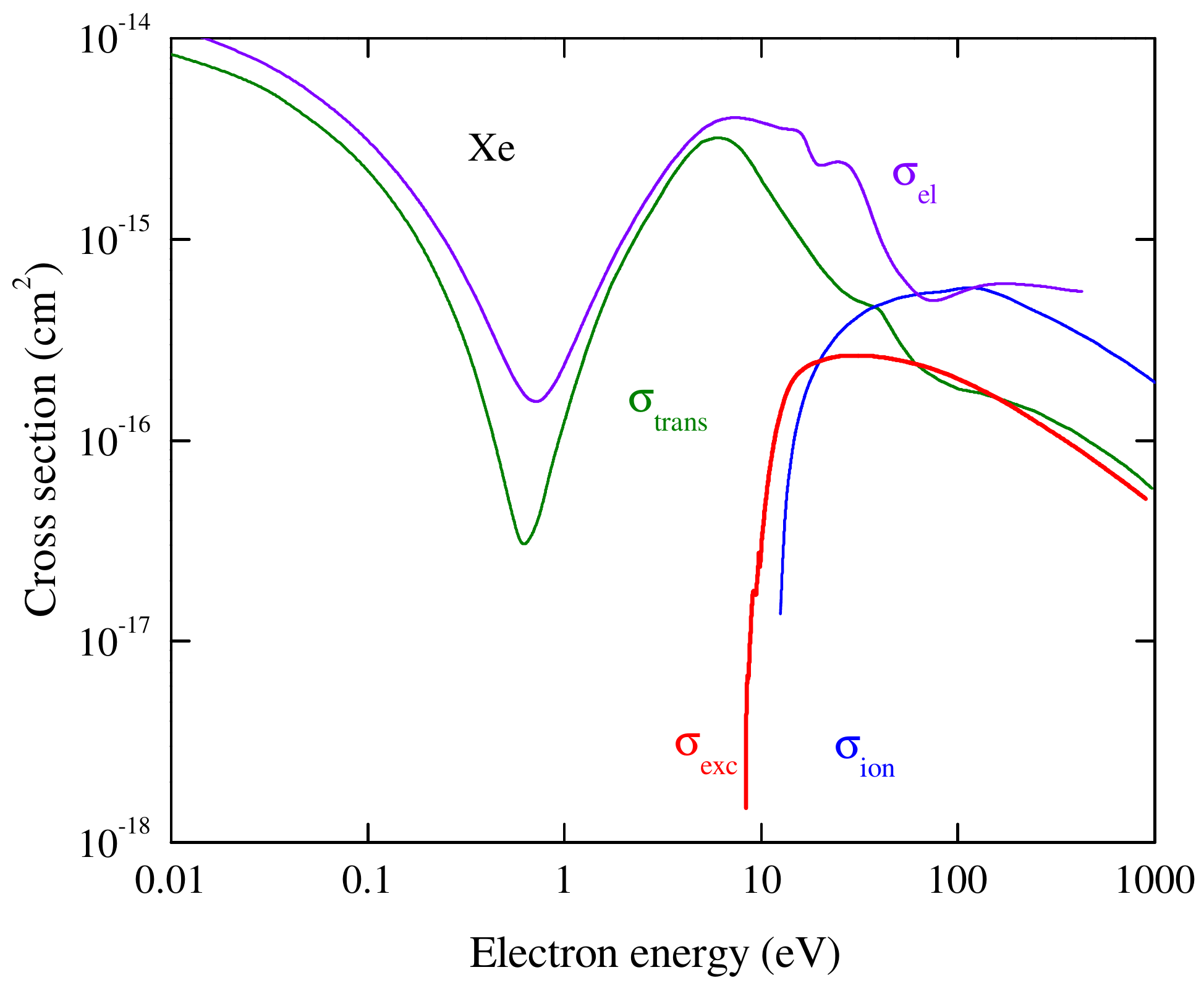}}\\
	\end{minipage}
	\hfill

	\caption{Electron scattering cross section in noble gases as a function of electron energy used in this work, obtained from the data bases~\cite{DBBiagi,DBBSR}: that of total elastic ($\sigma_{el}$), momentum transfer ($\sigma_{trans}$), excitation ($\sigma_{exc}$) and ionization ($\sigma_{ion}$). The excitation cross section was obtained as the sum of those for all given excitation states.}
	\label{Fig01}
\end{figure*}

\begin{figure*}[!h]
	\begin{minipage}[h]{0.29\linewidth}
		\center{\includegraphics[width=1\linewidth]{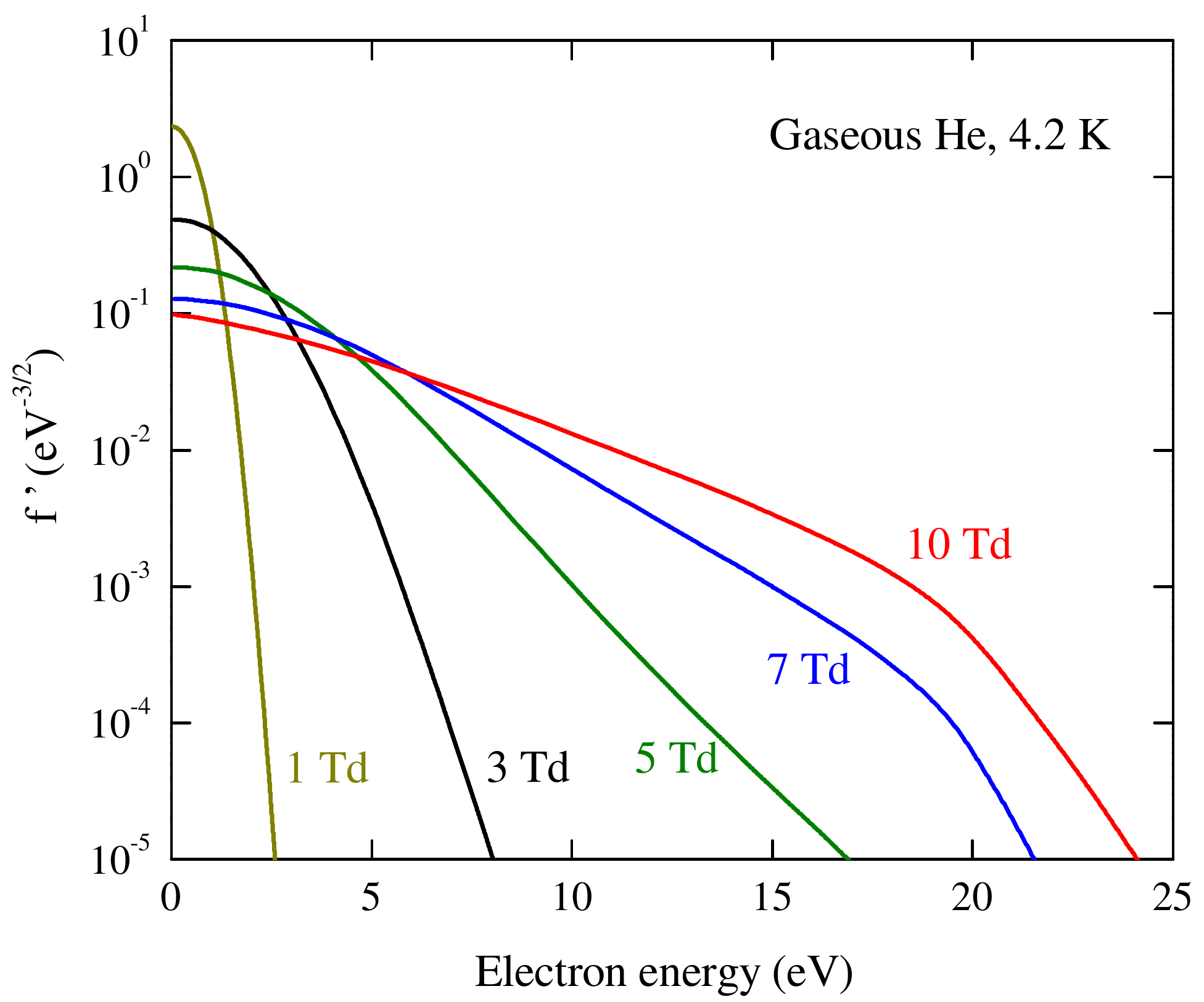}} \\
	\end{minipage}
	\hfill
	\begin{minipage}[h]{0.29\linewidth}
		\center{\includegraphics[width=1\linewidth]{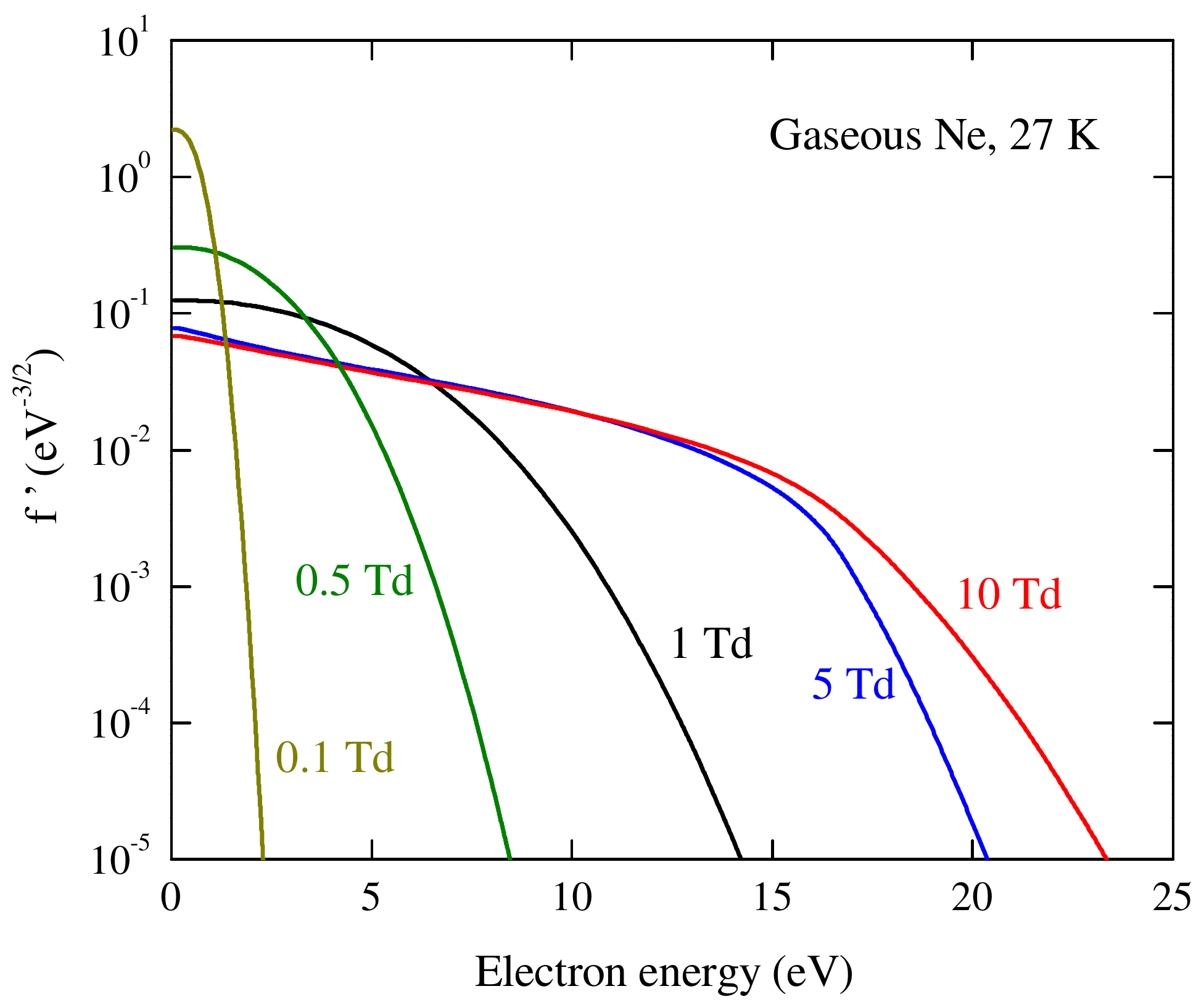}}
	\end{minipage}
	\vfill
	\begin{minipage}[h]{0.29\linewidth}
		\center{\includegraphics[width=1\linewidth]{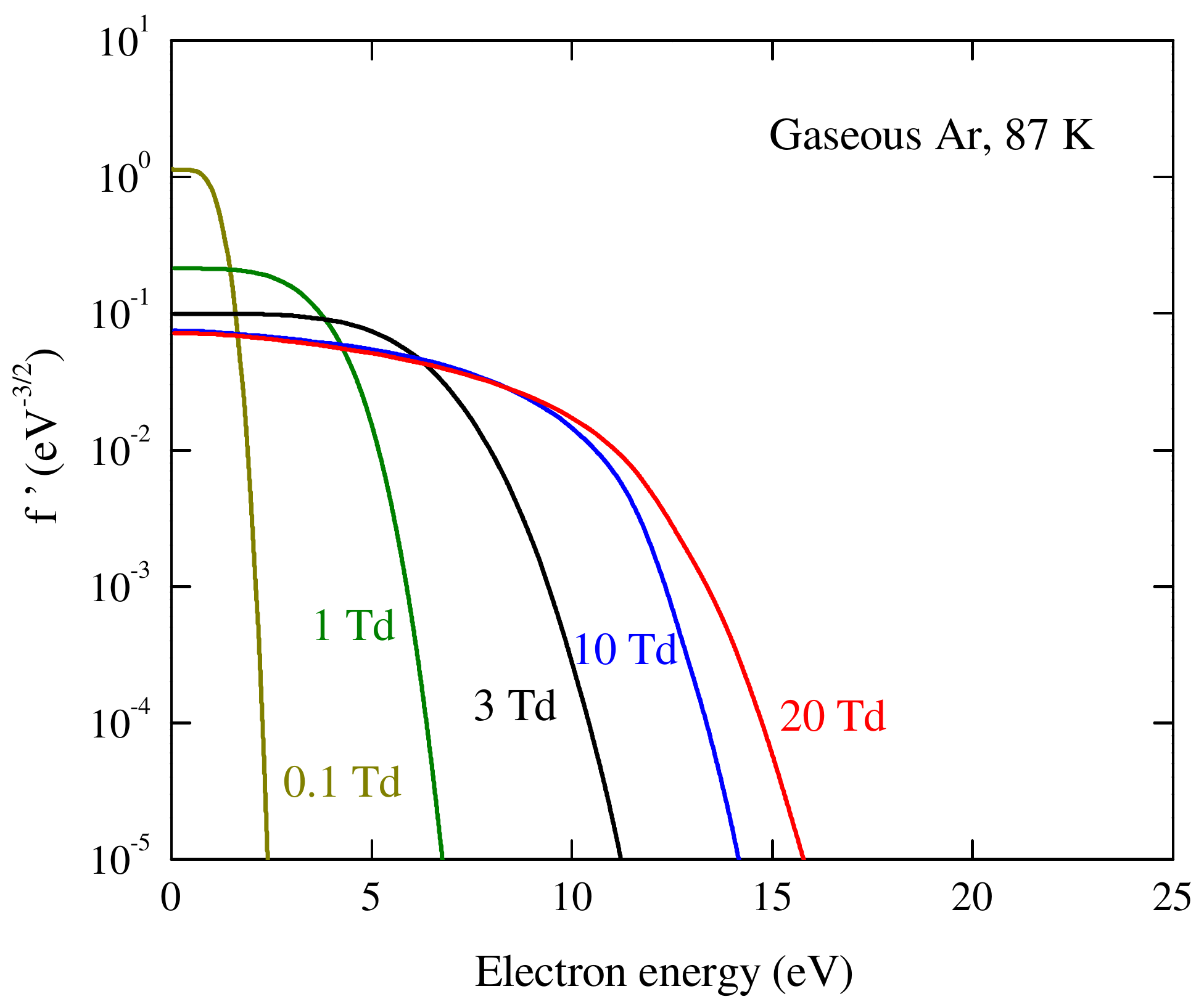}}\\
	\end{minipage}
	\hfill
	\begin{minipage}[h]{0.29\linewidth}
		\center{\includegraphics[width=1\linewidth]{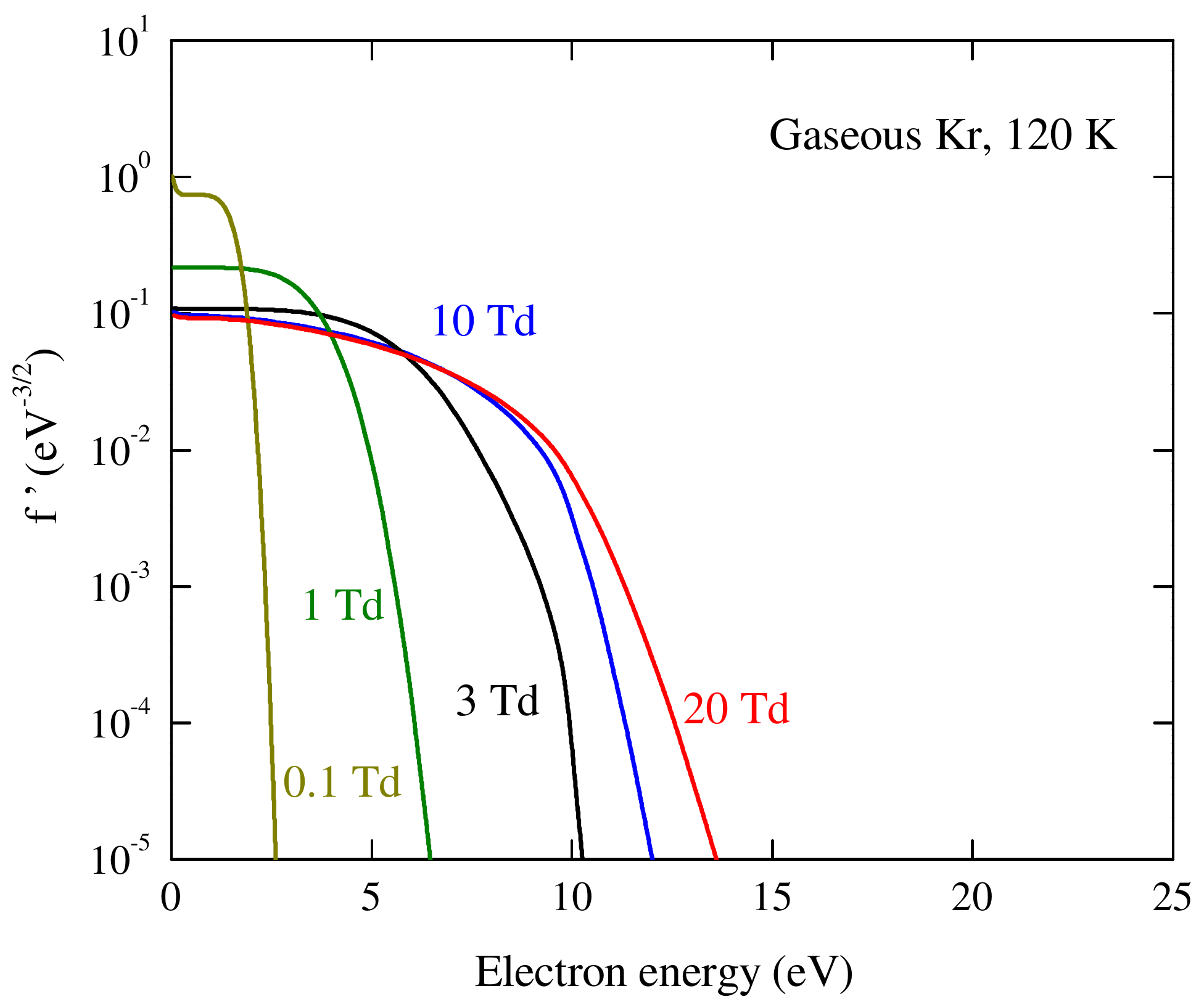}}\\
	\end{minipage}
	\hfill
	\begin{minipage}[h]{0.29\linewidth}
		\center{\includegraphics[width=1\linewidth]{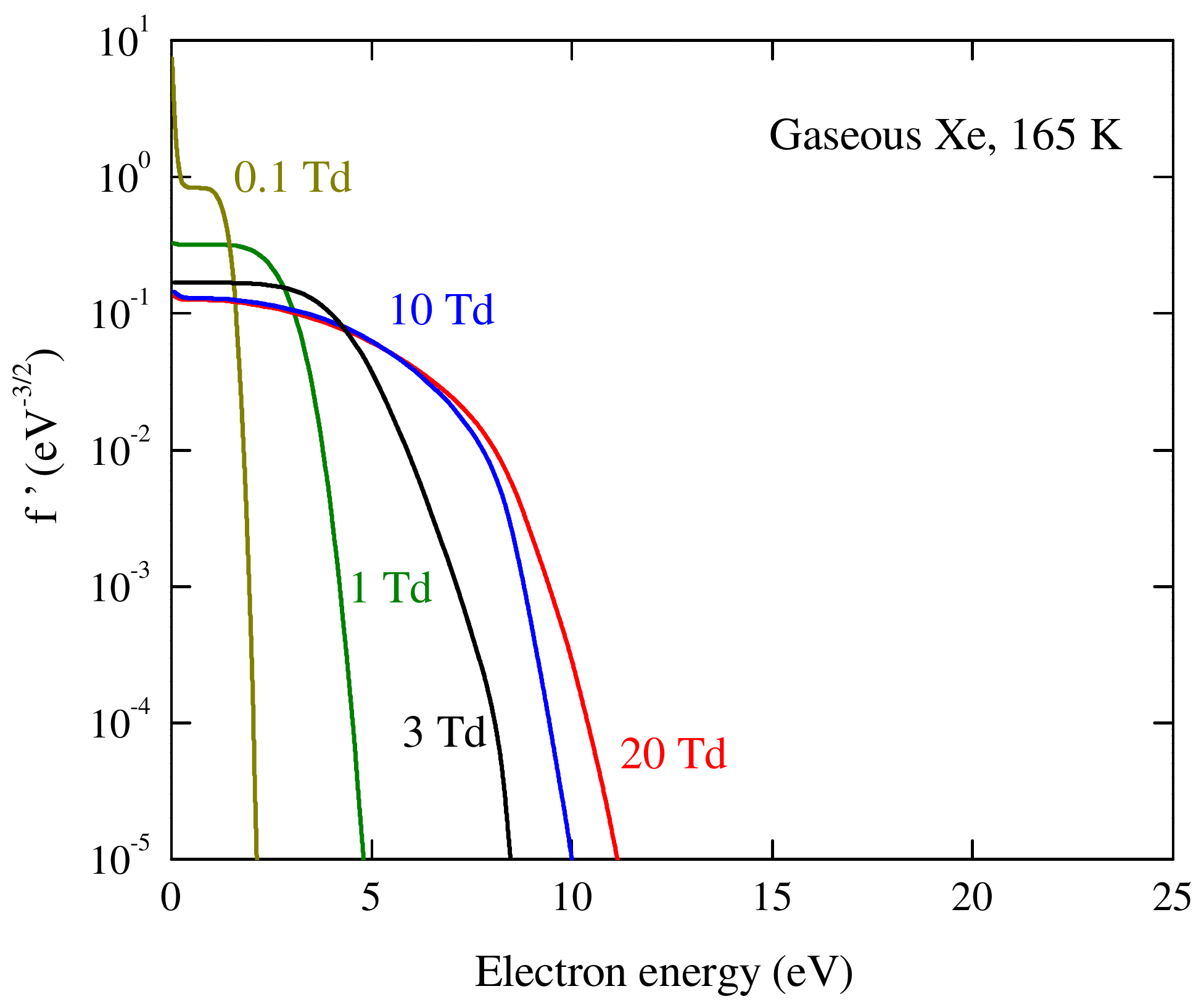}}\\
	\end{minipage}
	\hfill
	
	\caption{Electron energy distribution functions with a prime ($f^\prime$) in noble gases, normalized as in Eq.~\ref{Eq-norm-fprime}, calculated using Boltzmann equation solver BOLSIG+~\cite{Bolsig2}, at different reduced electric fields.}
	\label{Fig02}
\end{figure*}


\begin{figure}[h!]
		\center{\includegraphics[width=0.8\columnwidth]{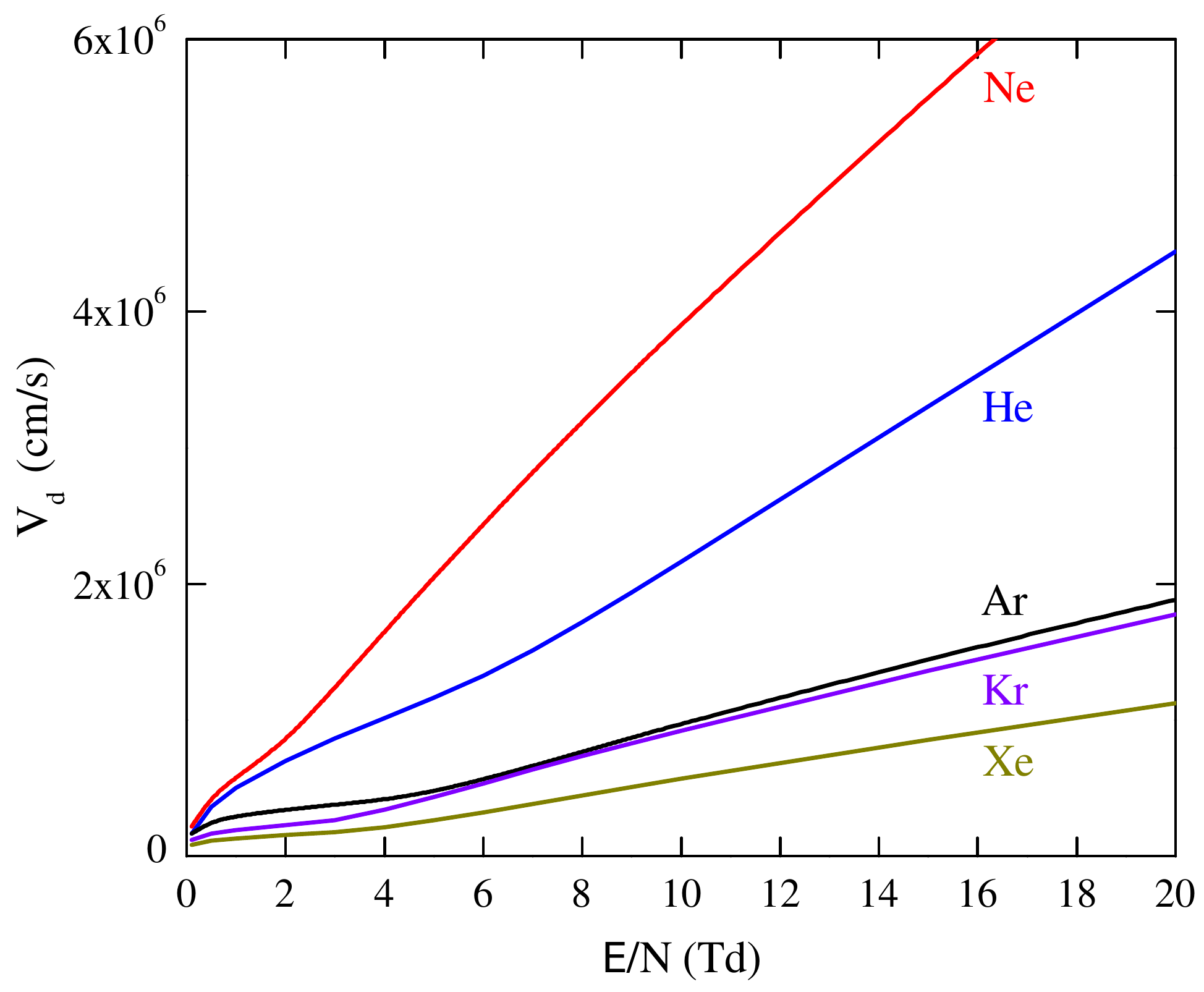}} \\
		\center{\includegraphics[width=0.8\columnwidth]{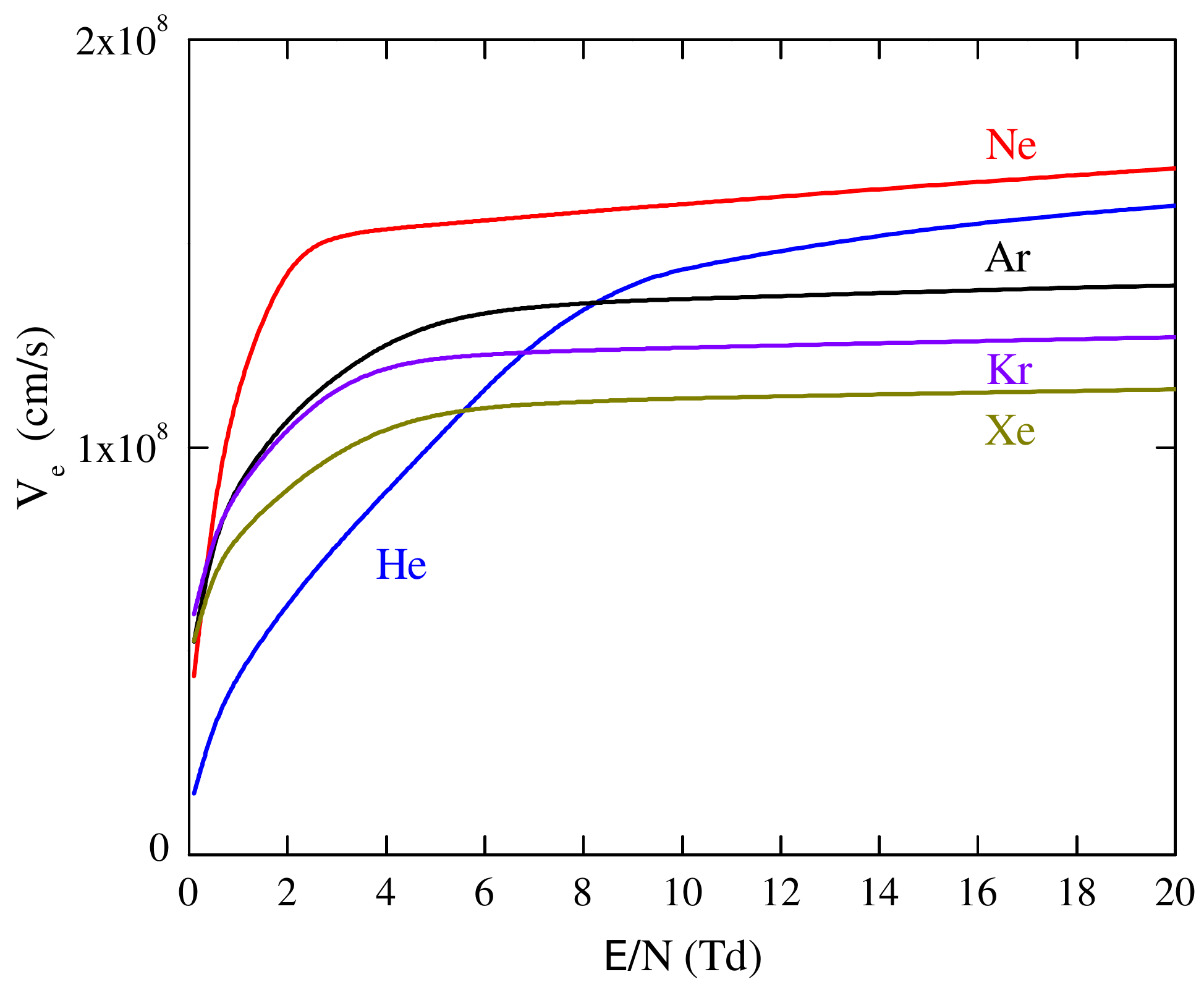}}
	\caption{Electron drift velocity (top) in noble gases and that of chaotic motion (bottom) as a function of the reduced electric field, calculated in this work using Boltzmann equation solver.}
	\label{Fig03}
\end{figure}

\begin{figure}[h!]
		\center{\includegraphics[width=0.8\columnwidth]{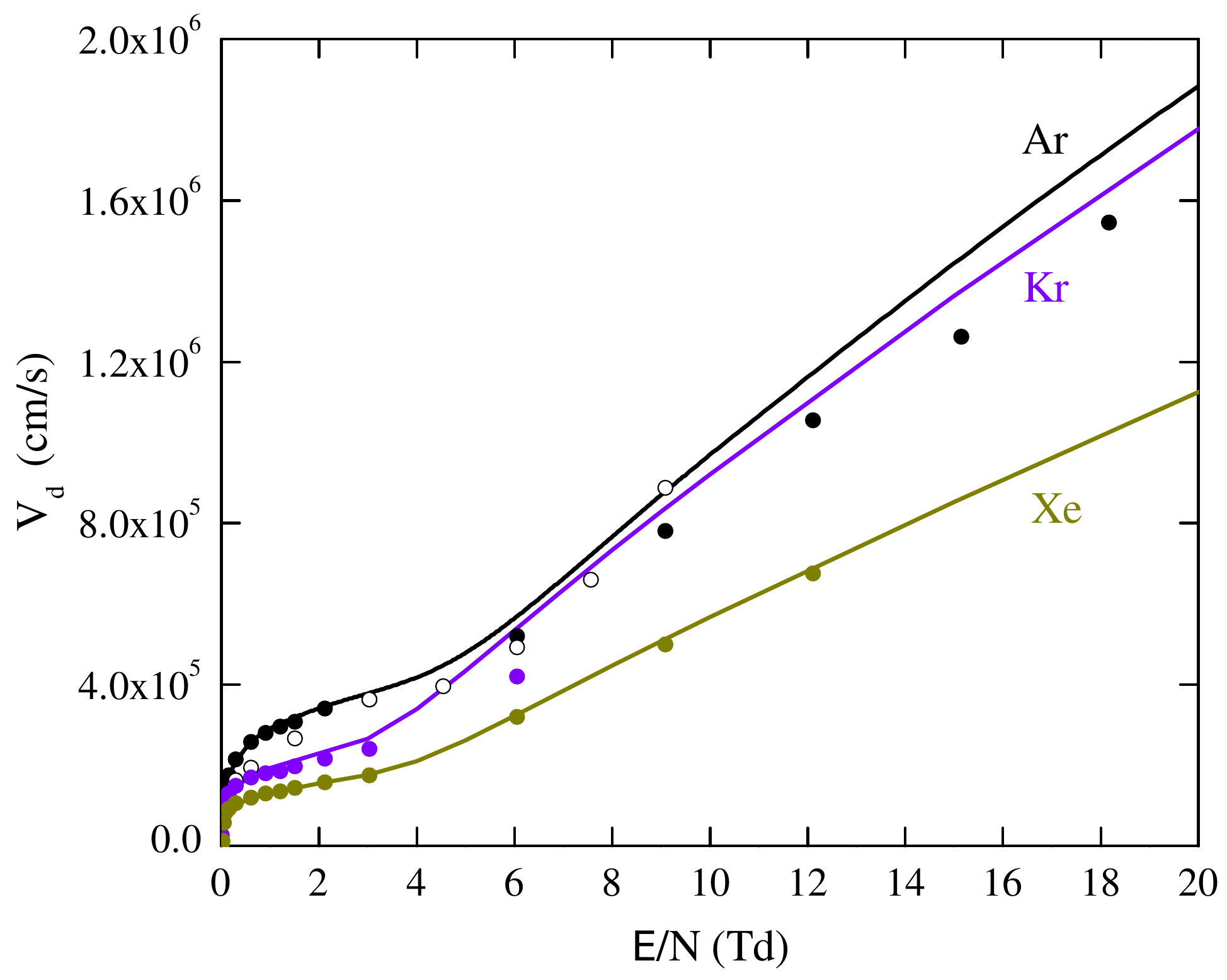}} \\
		\center{\includegraphics[width=0.8\columnwidth]{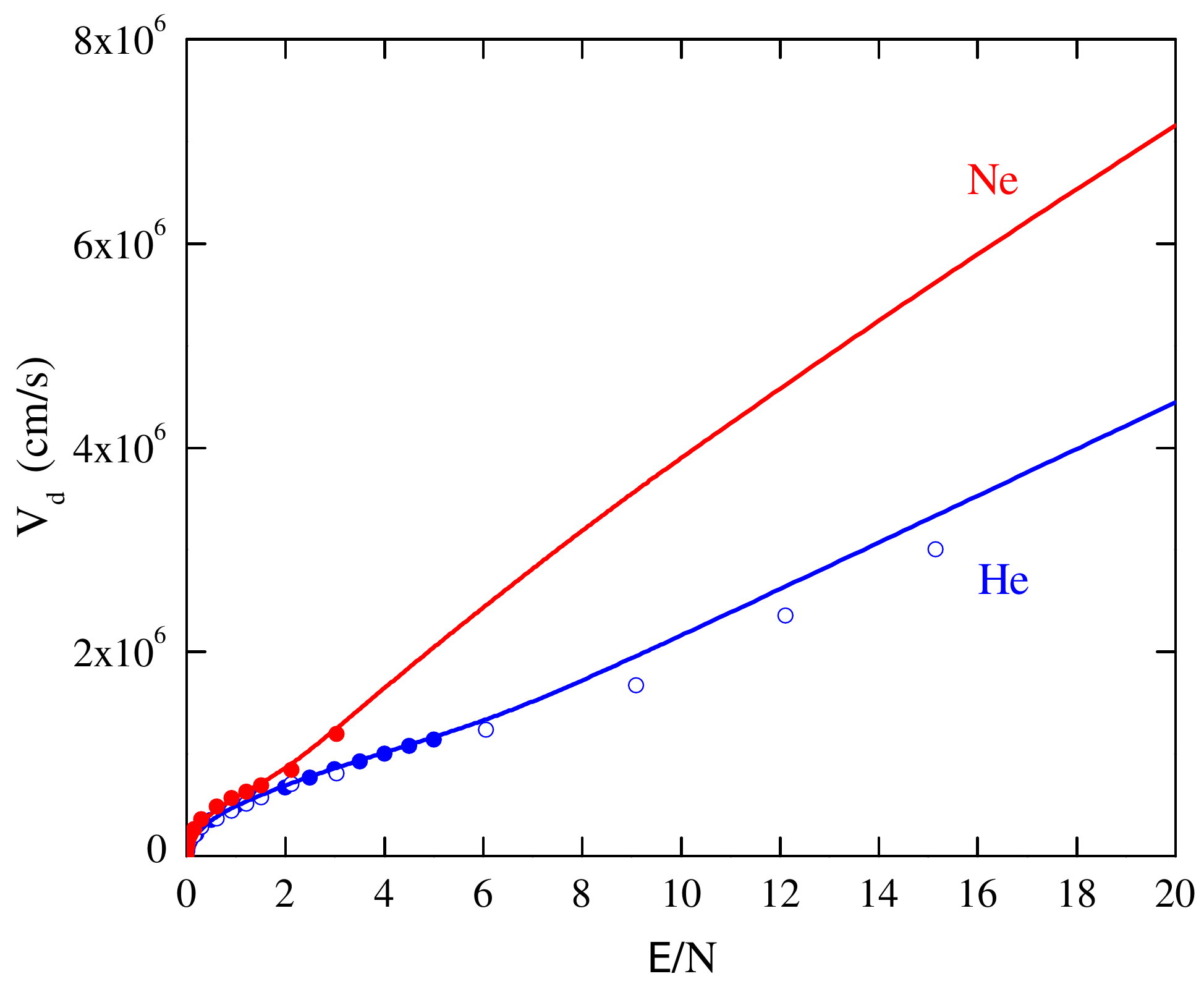}}
	\caption{Comparison of electron drift velocity ($\upsilon_d$) in noble gases theoretically calculated in this work (curves) with that measured in experiment \cite{Peisert84} (data points). The color of the curve and the data points is the same for a given gas.}
	\label{Fig031}
\end{figure}

\begin{figure}[h!]
		\center{\includegraphics[width=0.8\columnwidth]{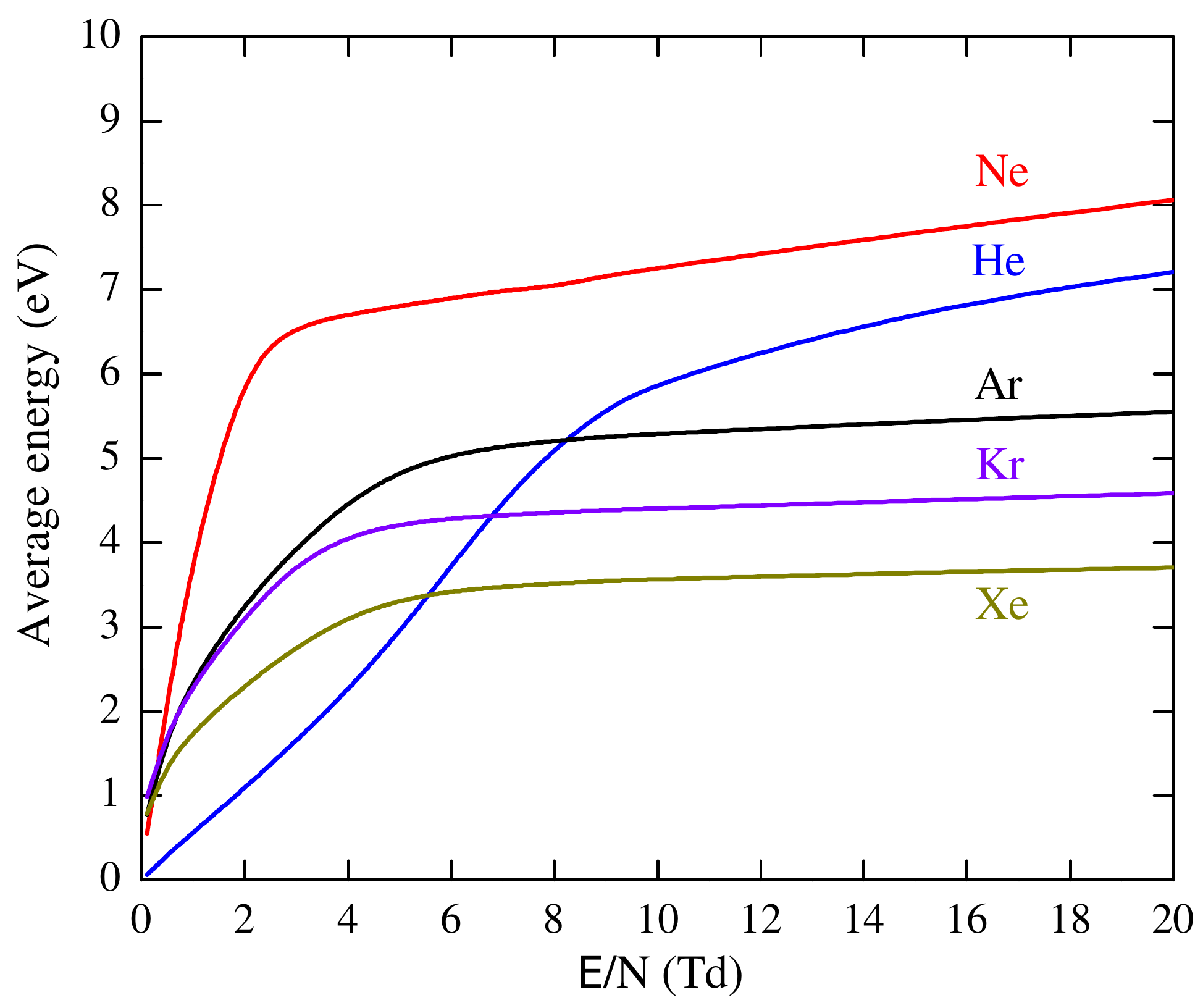}} \\
		\center{\includegraphics[width=0.8\columnwidth]{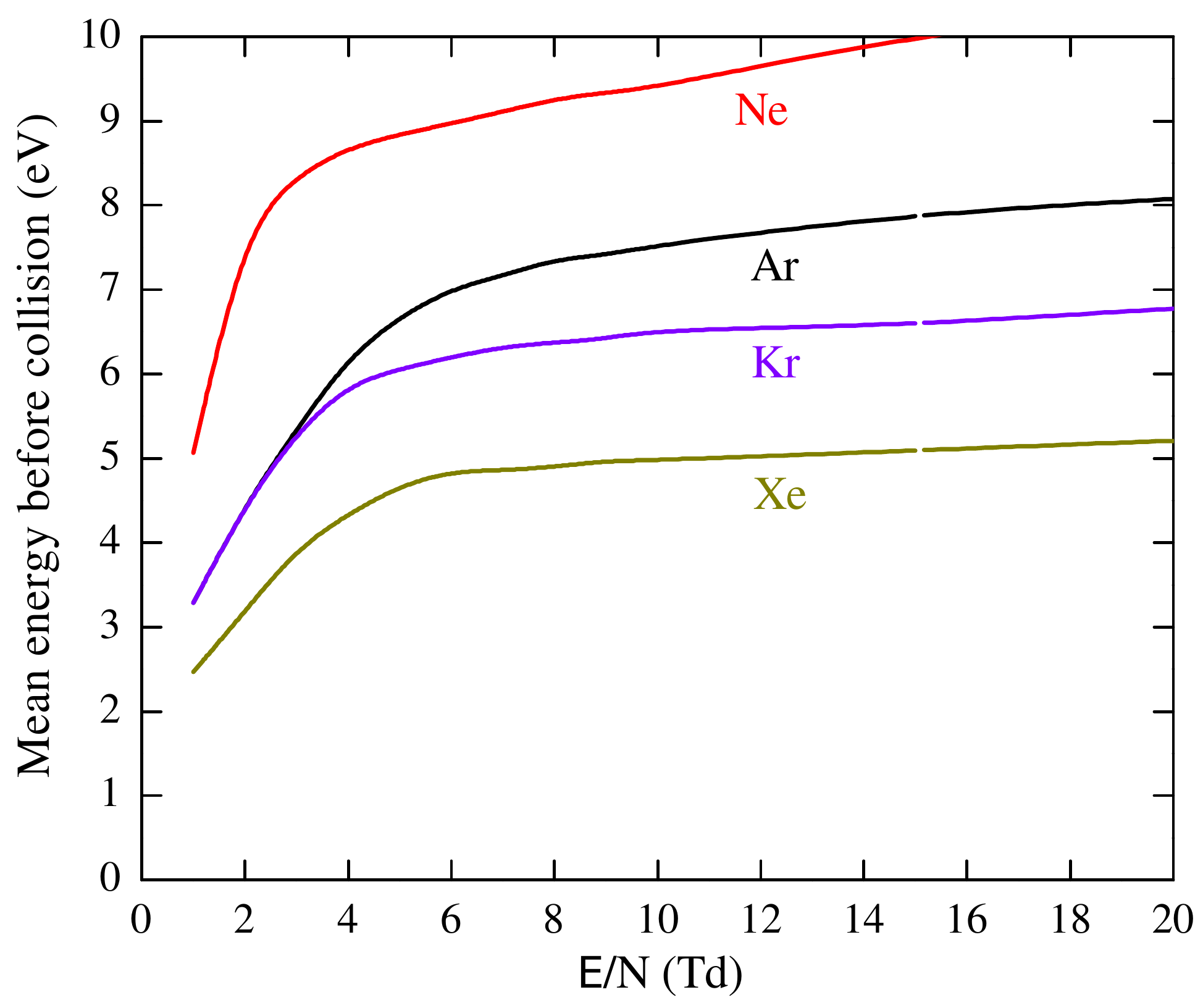}}
	\caption{Electron mean energy as a function of the reduced electric field, calculated in this work using Boltzmann equation solver (top). Also shown is the mean energy reached by the electrons before collisions, taken from~\cite{Oliveira11} where a microscopic approach was used for electron transport simulation (bottom).}
	\label{Fig04}
\end{figure}

In particular, Fig.~\ref{Fig01} presents the following cross sections for each noble gas: that of momentum-transfer and ionization taken from the Biagi database~\cite{DBBiagi}, that of excitation obtained as the sum of cross sections for all given excitation states taken from the Biagi database~\cite{DBBiagi}, and that of total elastic for Ne, Ar, Kr, and Xe taken from the BSR database~\cite{DBBSR}. 

It should be remarked that there are two opinions about what elastic cross section should appear in Eq.~\ref{Eq-sigma-el}: that of total elastic \cite{Firsov61,Kasyanov65,Biberman67,Kasyanov78} or momentum transfer (transport) \cite{Park00,Dalgarno66,Biberman67}.  
In the region of interest of electron energies (1-10 eV), where the contribution to NBrS cross section is maximal,
the transport cross section is slightly ($\leq$25\%) smaller than that
of total elastic in all noble gases, resulting is that the NBrS cross section would be appropriately reduced by this small factor (below 25\%), if to use the transport cross section instead of total elastic. This reduction is far smaller than the uncertainty of the theoretical model (factor of 2). Consequently, this could not change the conclusions of the calculations. Therefore in what follows, we use the total elastic cross section for all noble gases, except He where only the data on the transport cross section are available in the Biagi database.

Fig.~\ref{Fig02} shows examples of the calculated electron energy distribution function, namely the distribution functions with a prime ($f^\prime$) often used instead of $f$ and normalized as 
\begin{eqnarray}
\label{Eq-norm-fprime} 
\int\limits_{0}^{\infty} E^{1/2} f^\prime(E) \ dE = 1 \; .
\end{eqnarray}
$f^\prime$ is considered to be more enlightening than $f$, since in the limit of zero electric field it tends to Maxwellian distribution. 

Using the electron energy distribution functions, one can calculate the electron mean velocity of directed motion (drift velocity, $\upsilon_d$) and that of chaotic motion ($\upsilon_e$) for all noble gases: both are shown in Fig.~\ref{Fig03} as a function of the reduced electric field. It is possible to check the correctness of the distribution functions by comparing the calculated and measured electron drift velocities: this is done in Fig.~\ref{Fig031} using the experimental data compiled in \cite{Peisert84}. It can be seen that the theoretical and experimental drift velocities are in excellent agreement, thus confirming the correctness of the calculated distribution functions for all noble gases.

Electron chaotic velocity in Fig.~\ref{Fig03} (bottom) in fact reflects the dependence of the electron mean energy on the electric field: the mean energy was calculated in a similar way and is shown in Fig.~\ref{Fig04}. 

From Fig.~\ref{Fig02} one can see that at reduced electric fields exceeding 10~Td the distribution function tends to have a high-energy cut-off, corresponding to the onset of significant electron energy losses due to exciting the lowest atomic excitation level. The energy of the latter is lower for heavier noble gases (see Table~\ref{tbl:table1}, item 5) resulting in a correspondingly lower cut-off energy. 

This effect also explains why at higher electric fields the electron drift velocity, chaotic velocity and energy are lower for a heavier noble gas (see Figs.~\ref{Fig03} and \ref{Fig04}): these quantities decrease in the series Ne, Ar, Kr and Xe. 

It is interesting that He violates this rule: the electron drift velocity is always lower in He than in Ne, and the electron energy in He is lower than that of all other noble gases at reduced electric fields below 5 Td. This strangeness of He is fully explained by the properties of its electron elastic cross section which defines the electron transport in this energy range: see Fig.~\ref{Fig01}. Indeed, at electron energies below 3 eV, the electron elastic cross section in He is significantly larger than that of other noble gases, in particular due to the fact that it does not have a Ramsauer minimum in the elastic cross section, in contrast to Ar, Kr and Xe. The larger the elastic cross section, the lower the electron mean energy and velocity in this energy region.

It should be remarked that the distribution function and other quantities in Figs.~\ref{Fig02},~\ref{Fig03} and~\ref{Fig04} (top), obtained by solving the Boltzmann equation, are actually averaged over the collisions and hence also over the time between the collisions. On the other hand, in microscopic approach \cite{Oliveira11},  the energy just before the collision is used, which obviously exceeds the time-averaged energy since the electrons are accelerated by the electric field between the collisions: compare fig.~\ref{Fig04} (top) and fig.~\ref{Fig04} (bottom). In microscopic approach this energy is substituted into the formulas for the cross section, when simulating the event. This raises the question which distribution function should be used in equations~\ref{Eq-NBrS-el-yield}, \ref{Eq-NBrS-el-yield-spectrum} and \ref{Eq-ord-el-yield}: that of time-averaged, obtained by solving the Boltzmann equation, or that averaged before collisions, obtained in microscopic approach? 

In our previous work \cite{Buzulutskov18} both distributions functions were used, and the variation of the results between them was taken to determine the theoretical uncertainty: the latter reached a factor of 2 for the EL yield. Such a theoretical uncertainty can be considered as quite acceptable for the EL yields varying by 4 orders of magnitude (see Fig.~\ref{Fig00Ar}). For simplicity, in this work we restrict ourselves to only one type of distribution function, namely to that obtained by solving the Boltzmann equation.

\begin{figure*}[!hp]
	\begin{minipage}[h]{0.49\linewidth}
		\center{\includegraphics[width=1\linewidth]{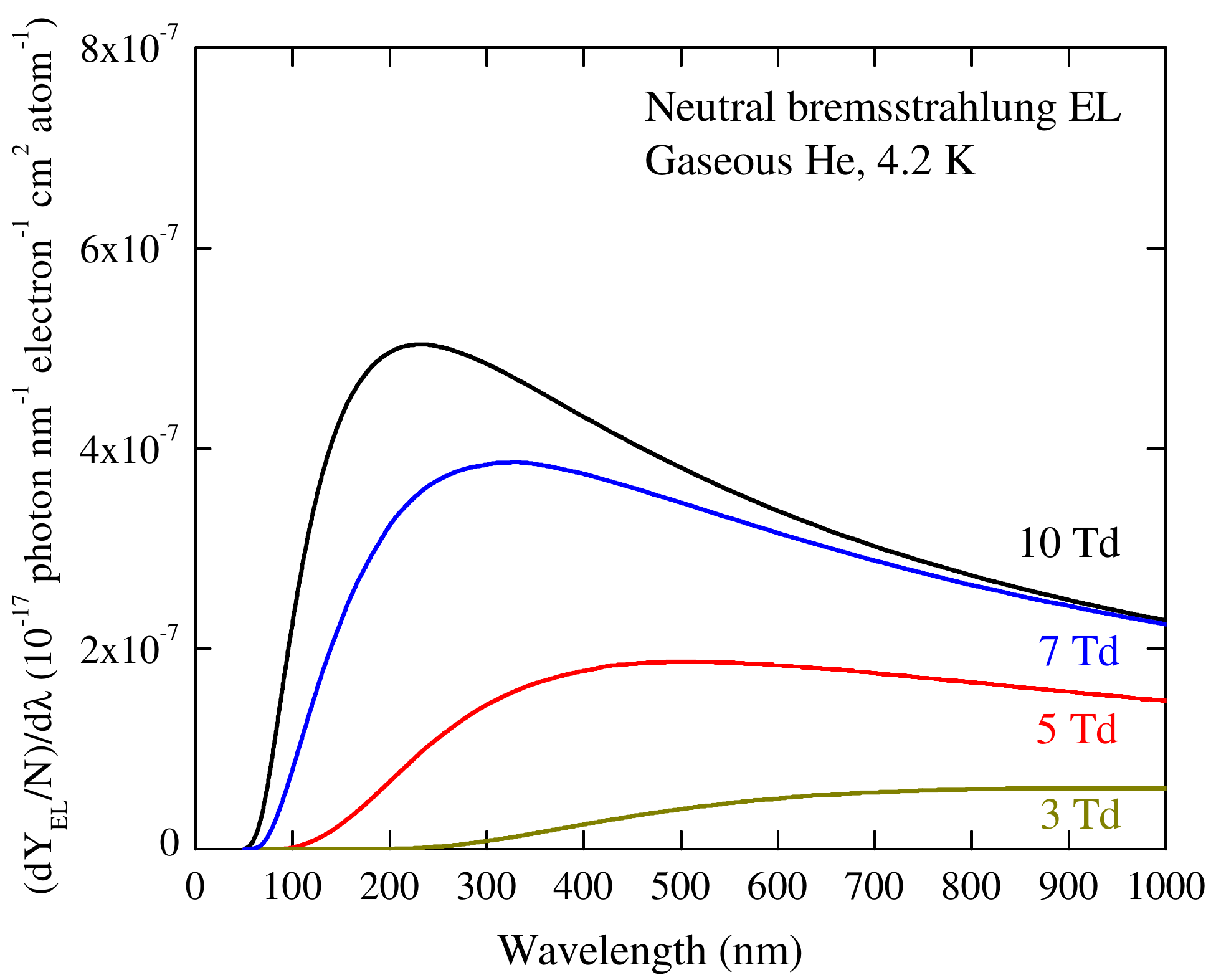}} \\
	\end{minipage}
	\hfill
	\begin{minipage}[h]{0.49\linewidth}
		\center{\includegraphics[width=1\linewidth]{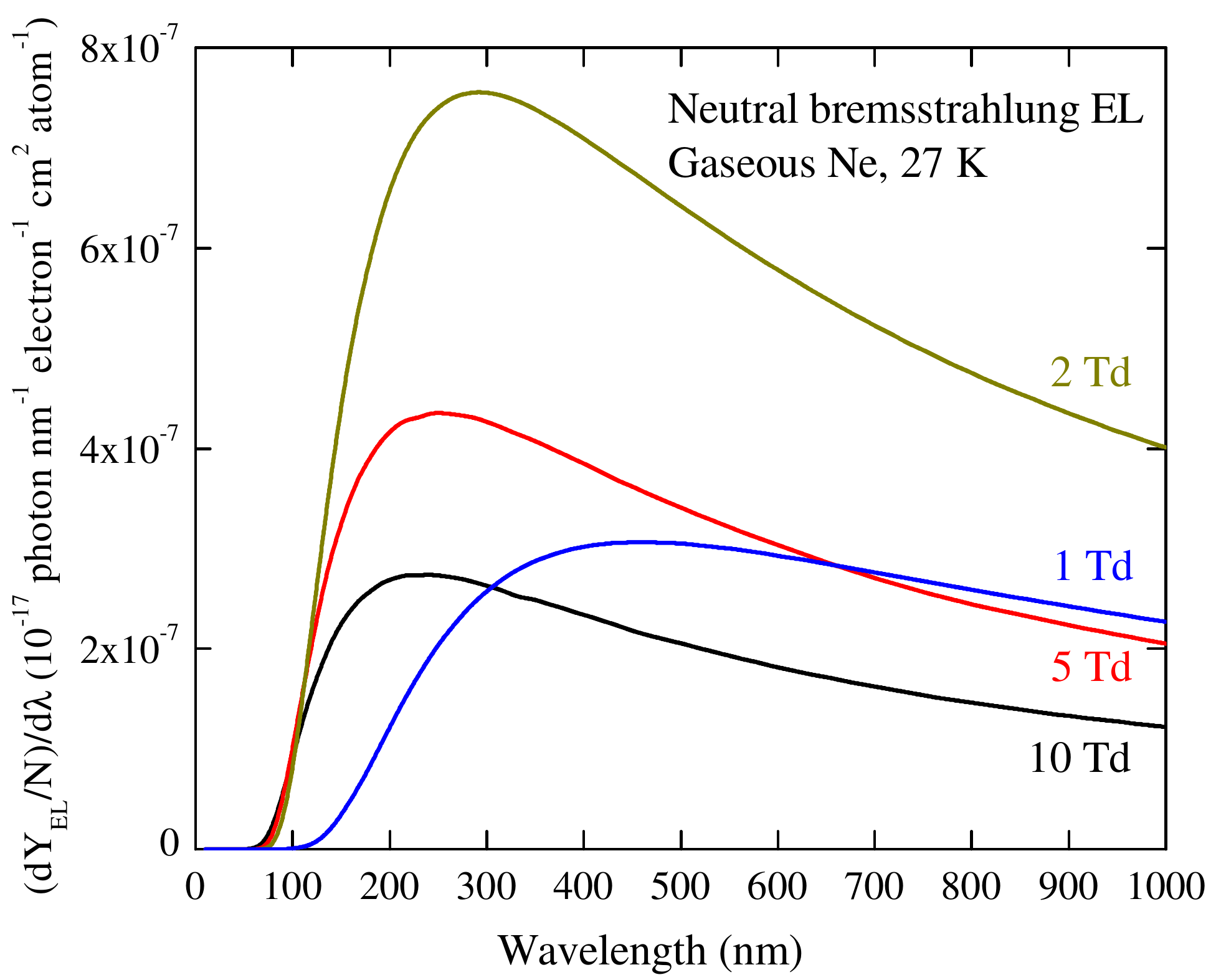}}
	\end{minipage}
	\vfill
	\begin{minipage}[h]{0.49\linewidth}
		\center{\includegraphics[width=1\linewidth]{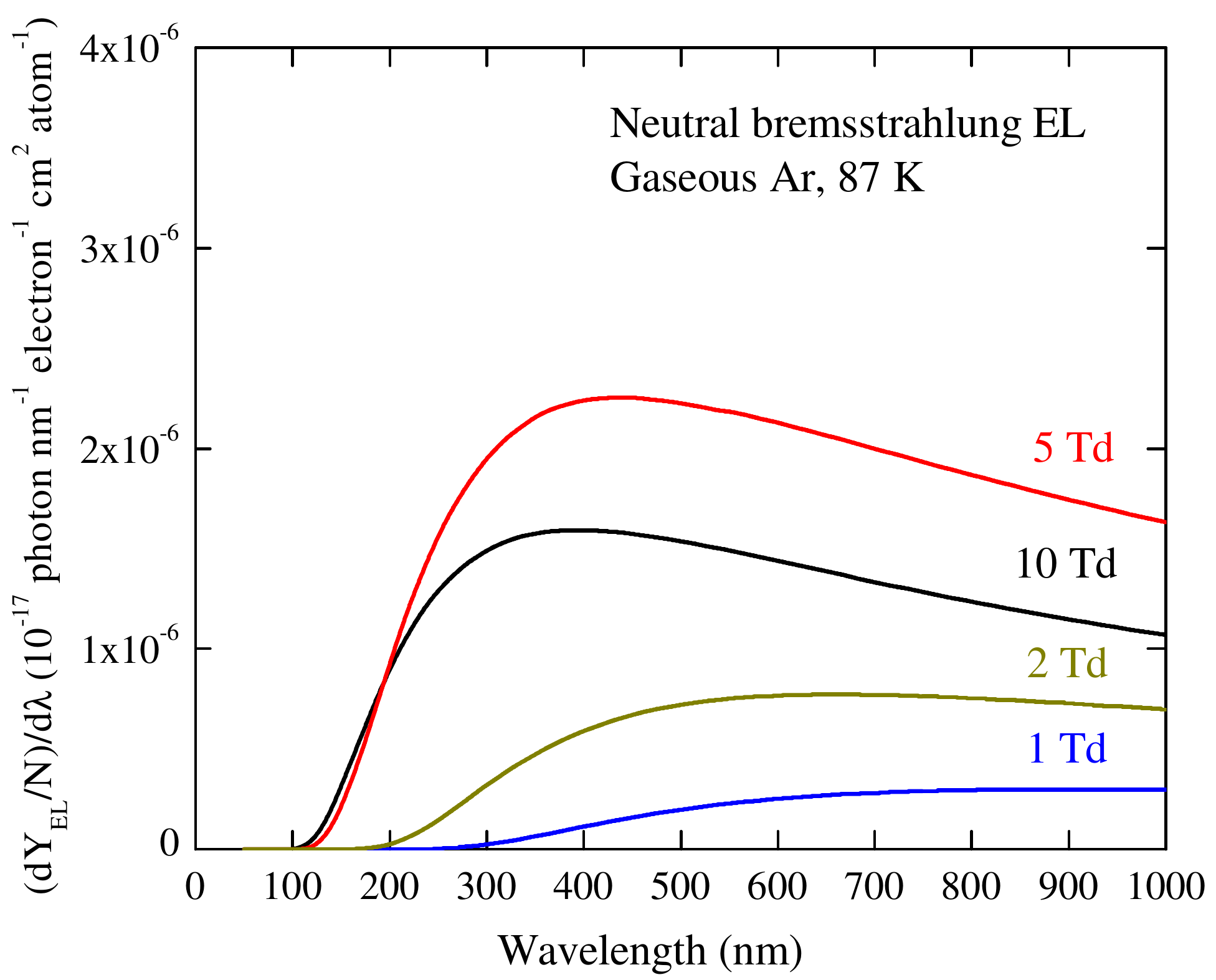}}\\
	\end{minipage}
	\hfill
	\begin{minipage}[h]{0.49\linewidth}
		\center{\includegraphics[width=1\linewidth]{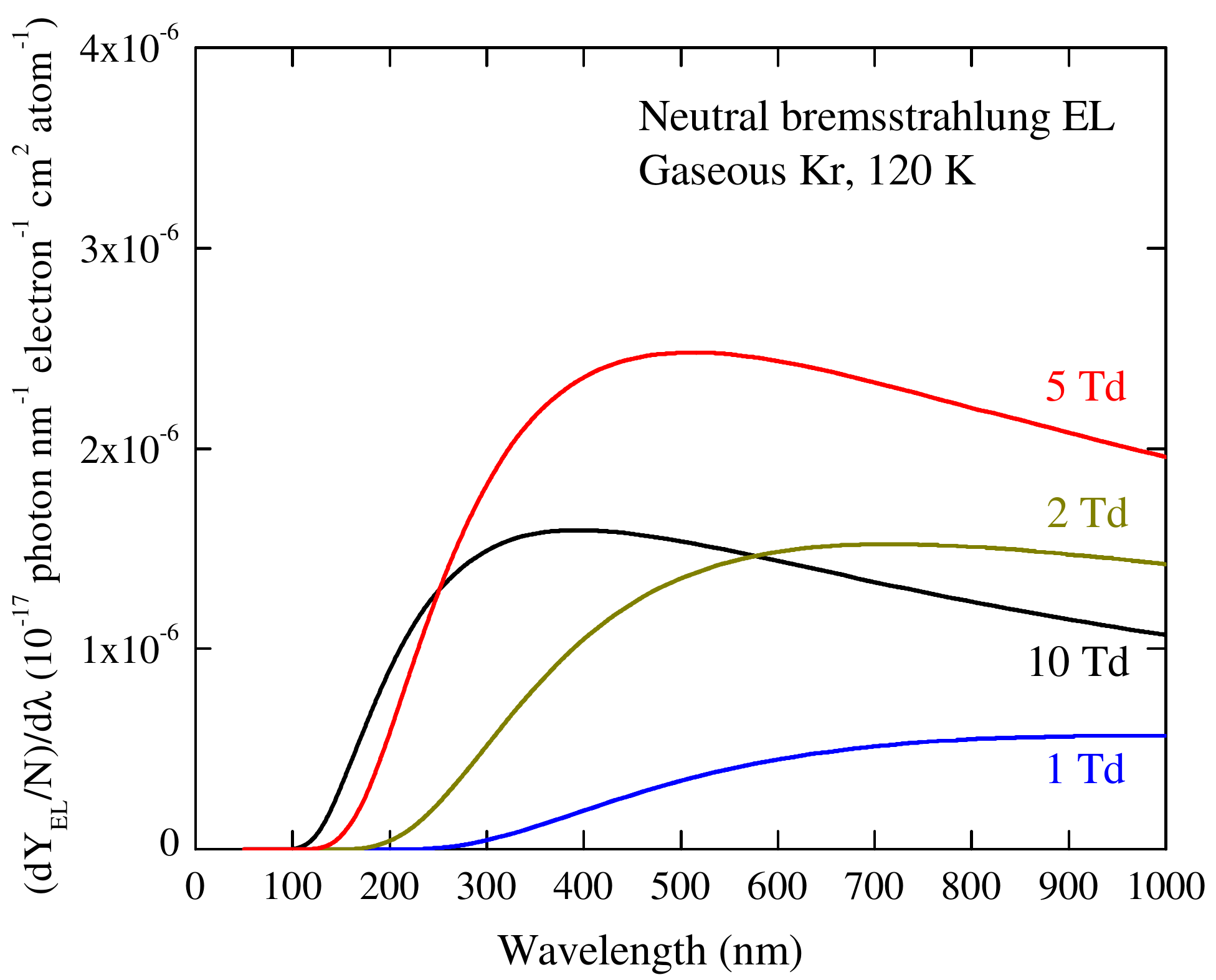}}\\
	\end{minipage}
	\begin{minipage}[h]{0.49\linewidth}
	\center{\includegraphics[width=1\linewidth]{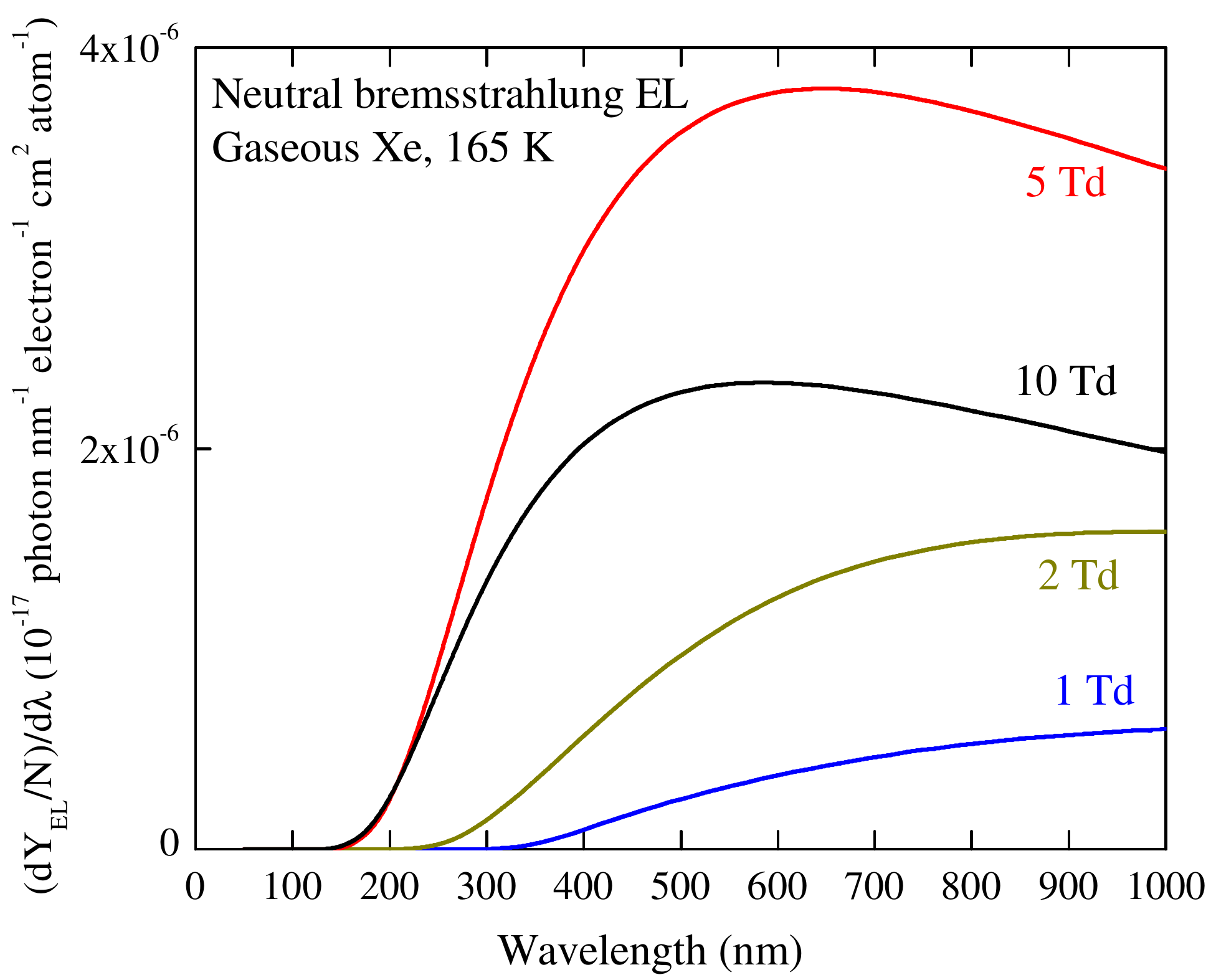}}\\
\end{minipage}
	\vfill
	
	\caption{Spectra of NBrS EL yield in noble gases at different reduced electric fields, calculated using Eq.~\ref{Eq-NBrS-el-yield-spectrum}.}
	\label{Fig07}
\end{figure*}

\begin{figure}[!h]
	\centering
	\includegraphics[width=1\linewidth]{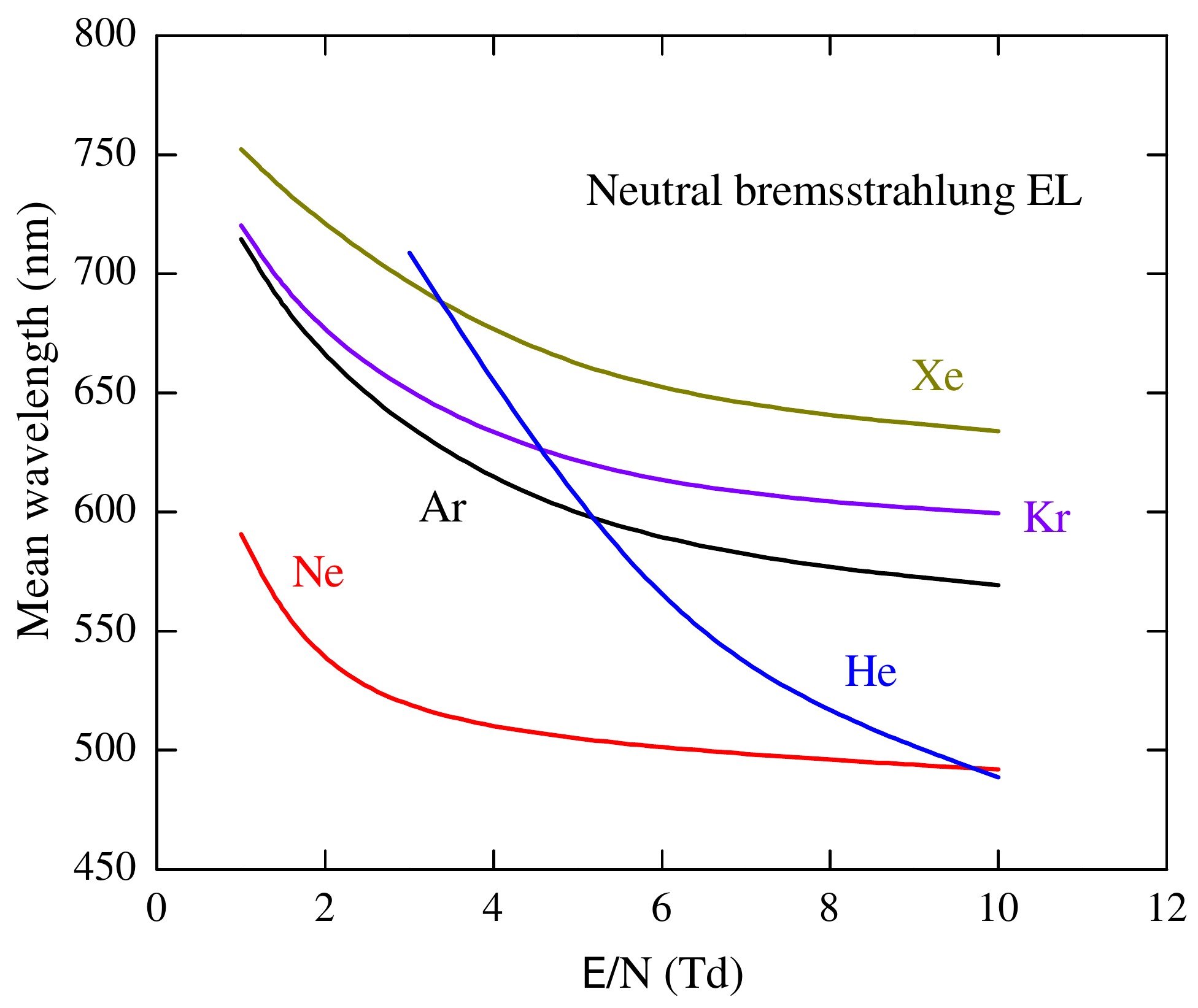}
	\caption{Mean wavelength of NBrS EL below 1000 nm in noble gases as a function of the reduced electric field.}
	\label{Fig07a}
\end{figure}

\begin{table*} [h!]
	\caption{Properties of noble gases, and parameters of neutral bremsstrahlung (NBrS) and excimer electroluminescence (EL) theoretically calculated in this work.}
	\label{tbl:table1}
	\begin{center}
		\begin{tabular}{p{0.2cm}p{4.0cm}p{1.6cm}p{2.2cm}p{2.2cm}p{2.2cm}p{2.3cm}}
			No. & Parameter & He & Ne & Ar & Kr & Xe \\
			\\
			(1) & Boiling temperature at 1.0~atm, $T_b$~\cite{Fastovsky71} (K)  & $4.215$ & $27.07$ & $87.29$ & $119.80$ & $165.05$ \\
			\\
			(2) & Gas atomic density at $T_b$ and 1.0 atm, derived from~\cite{Fastovsky71} (cm$^{-3}$) & $2.37\cdot10^{21}$ & $3.41\cdot10^{20}$ & $8.62\cdot10^{19}$ & $6.18\cdot10^{19}$ & $5.75\cdot10^{19}$ \\
			\\
			(3) & Liquid atomic density at $T_b$ and 1.0 atm, derived from~\cite{Fastovsky71} (from ~\cite{Theeuwes70} for Xe) (cm$^{-3}$) & $1.89\cdot10^{22}$ & $3.59\cdot10^{22}$ & $2.10\cdot10^{22}$ & $1.73\cdot10^{22}$ & $1.35\cdot10^{22}$ \\
			\\
			(4) & Sub-excitation Feshbach resonances description  and their energies~\cite{Schulz73} (eV) &  \scriptsize He$^-(1s2s^2\ ^2S)$ \footnotesize $19.34$ & \scriptsize Ne$^-(2p^53s^2\ ^2P_{3/2})$ \footnotesize $16.11$ \scriptsize Ne$^-(2p^53s^2\ ^2P_{1/2})$ \footnotesize $16.20$& \scriptsize Ar$^-(3p^54s^2\ ^2P_{3/2})$ \footnotesize $11.11$ \scriptsize Ar$^-(3p^54s^2\ ^2P_{1/2})$ \footnotesize $11.28$ & \scriptsize Kr$^-(4p^55s^2\ ^2P_{3/2})$ \footnotesize $9.51$ \scriptsize Kr$^-(4p^55s^2\ ^2P_{1/2})$ \footnotesize $10.17$& \scriptsize Xe$^-(5s^56s^2\ ^2P_{3/2})$ \footnotesize $7.90$ \scriptsize Xe$^-(5s^56s^2\ ^2P_{1/2})$ \footnotesize  $8.48$\\
			\\
			(5) & Lowest excitation level and its energy ~\cite{DBBiagi} (eV) & \scriptsize He$^*(1s2s\,^3S)$ \footnotesize $19.82$ & \scriptsize Ne$^*(2p^53s\,^3P_2)$ \footnotesize $16.62$& \scriptsize Ar$^*(3p^54s\,^3P_2)$ \footnotesize $11.55$ & \scriptsize Kr$^*(4p^55s\,^3P_2)$ \footnotesize $9.92$ & \scriptsize Xe$^*(5s^56s\,^3P_2)$ \footnotesize $8.32$\\
			\\
			(6) & Reduced electric field for NBrS EL at which EL yield is maximum, $\mathcal{E}/N_{max}$ (Td) & $9.3$ & $2.3$& $5.5$ & $4.0$ & $4.8$\\
			\\
			(7) & Nominal threshold in reduced electric field for excimer EL (Td) & $6.0$ & $1.5$& $4.0$ & $3.0$ & $3.5$\\
			\\
			(8) & Number of photons for NBrS EL at $\mathcal{E}/N_{max}$, produced by drifting electron in 1~cm thick EL gap at $T_b$ and 1.0~atm & $7.75$ & $1.57$& $1.35$ & $1.15$ & $1.41$\\
			\\
			(9) & Number of photons for excimer EL at $\mathcal{E}/N_{max}$, produced by drifting electron in 1~cm thick EL gap at $T_b$ and 1.0~atm & $3.32\cdot10^{3}$ & $1.94\cdot10^{2}$& $1.15\cdot10^{2}$ & $8.2\cdot10^{1}$ & $1.15\cdot10^{2}$\\
			\\
			(10) & Electric field strength, corresponding to reduced electric field of 1~Td at $T_b$ and 1.0 atm (kV/cm) & $23.7$ & $3.41$ & $0.86$ & $0.62$ & $0.58$ \\ 
			\\
			(11) & Voltage needed to provide reduced electric field of 10~Td in 1~cm thick EL gap at $T_b$ and 1.0~atm (kV) & $237$ & $34.1$& $8.6$ & $6.2$ & $5.8$\\
			\\
		\end{tabular}
	\end{center}
\end{table*}

\section{EL spectra and yields}

Figs.~\ref{Fig07} show the NBrS spectra of the reduced EL yield. The spectra were calculated by numerical integration of Eq.~\ref{Eq-NBrS-el-yield-spectrum}. One can compare these to the spectra of excimer luminescence shown in Fig.~\ref{Int00}. 

One can see that the NBrS EL spectra are very similar in all noble gases: they are rather flat, extending from the UV to visible and NIR range. In each noble gas, the NBrS EL spectrum has a broad maximum that gradually moves to shorter wavelengths with the electric field. The fact that the NBrS EL spectra do not differ much in noble gases is confirmed in Fig.~\ref{Fig07a}, showing how the mean wavelength of the spectrum below 1000 nm depends on the electric field: it decreases from 700-750 nm at 1 Td to 550-650 nm at 10 Td, in Ar, Kr and Xe, and to 500 nm, in He and Ne. 

In all noble gases, the vast majority of the spectrum is above 200 nm  (in the UV, visible and NIR range), i.e. just in the sensitivity region of commonly used PMTs and SiPMs. This implies a possible practical application of NBrS EL, namely the method of direct optical readout of two-phase detectors in the visible range, i.e. without using WLS. Such a technique has been recently demonstrated in two-phase Ar detector with direct SiPM-matrix readout \cite{Aalseth21}. 

\begin{figure*}[!hp]
	\begin{minipage}[h]{0.49\linewidth}
		\center{\includegraphics[width=1\linewidth]{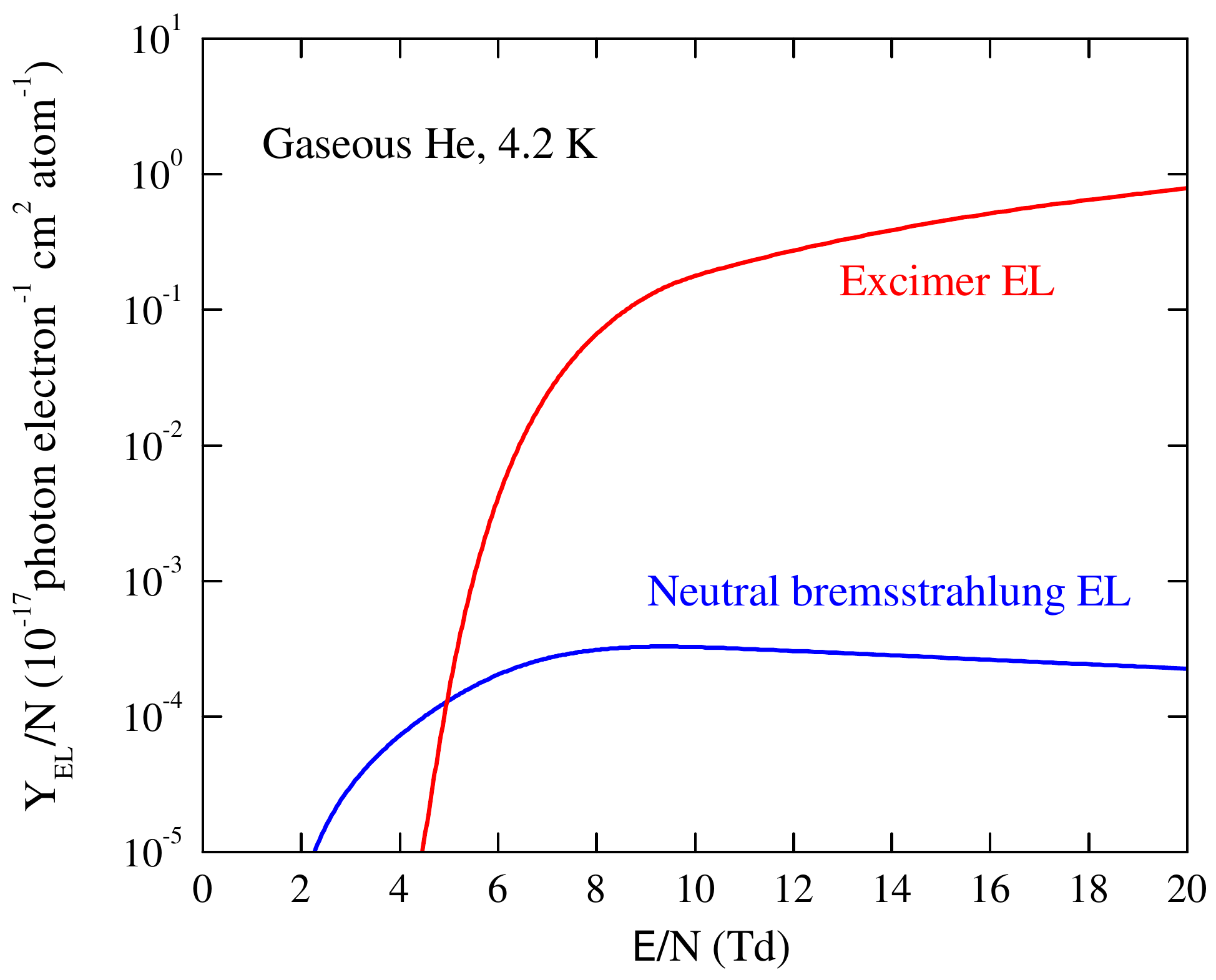}} \\
	\end{minipage}
	\hfill
	\begin{minipage}[h]{0.49\linewidth}
		\center{\includegraphics[width=1\linewidth]{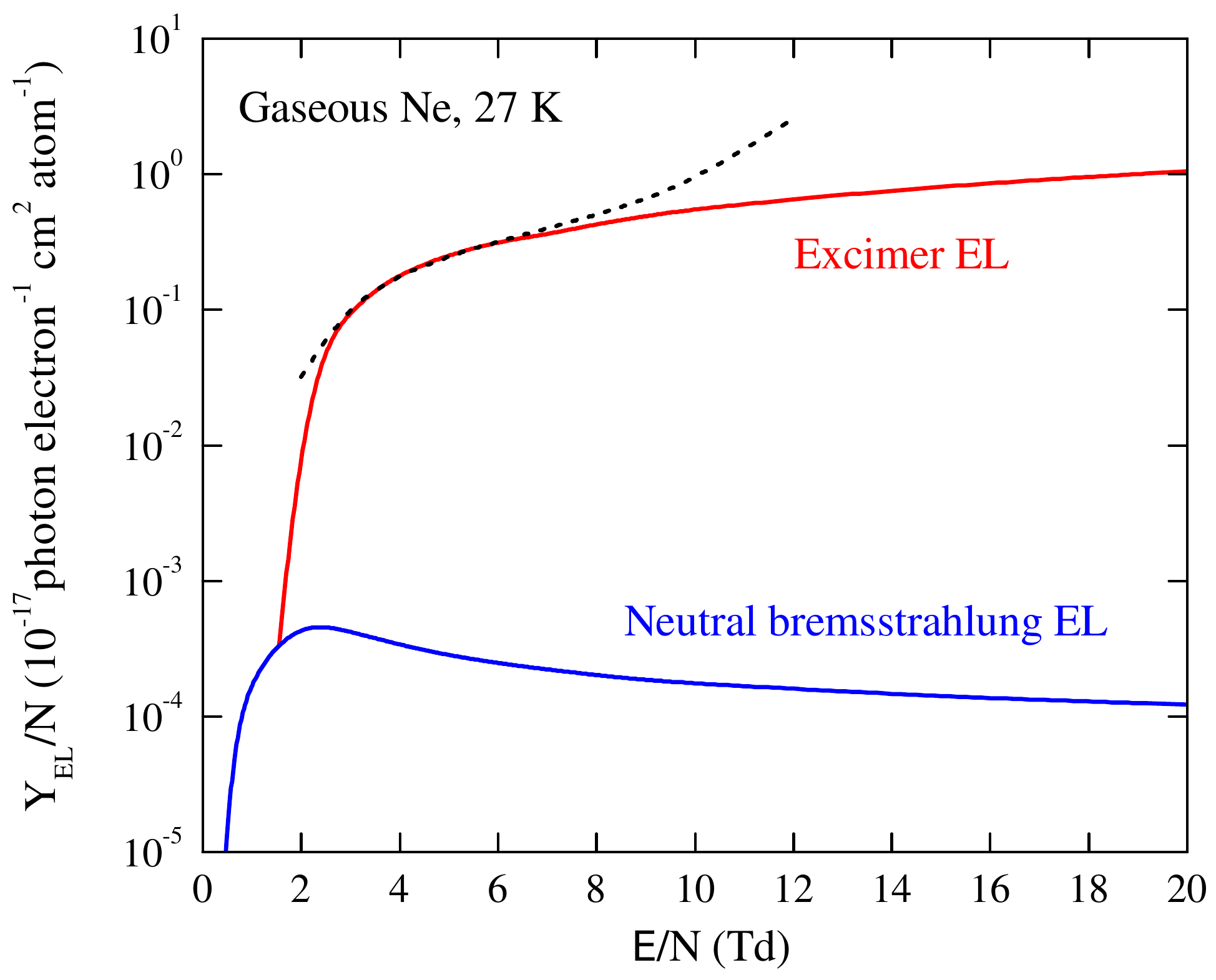}}
	\end{minipage}
	\vfill
	\begin{minipage}[h]{0.49\linewidth}
		\center{\includegraphics[width=1\linewidth]{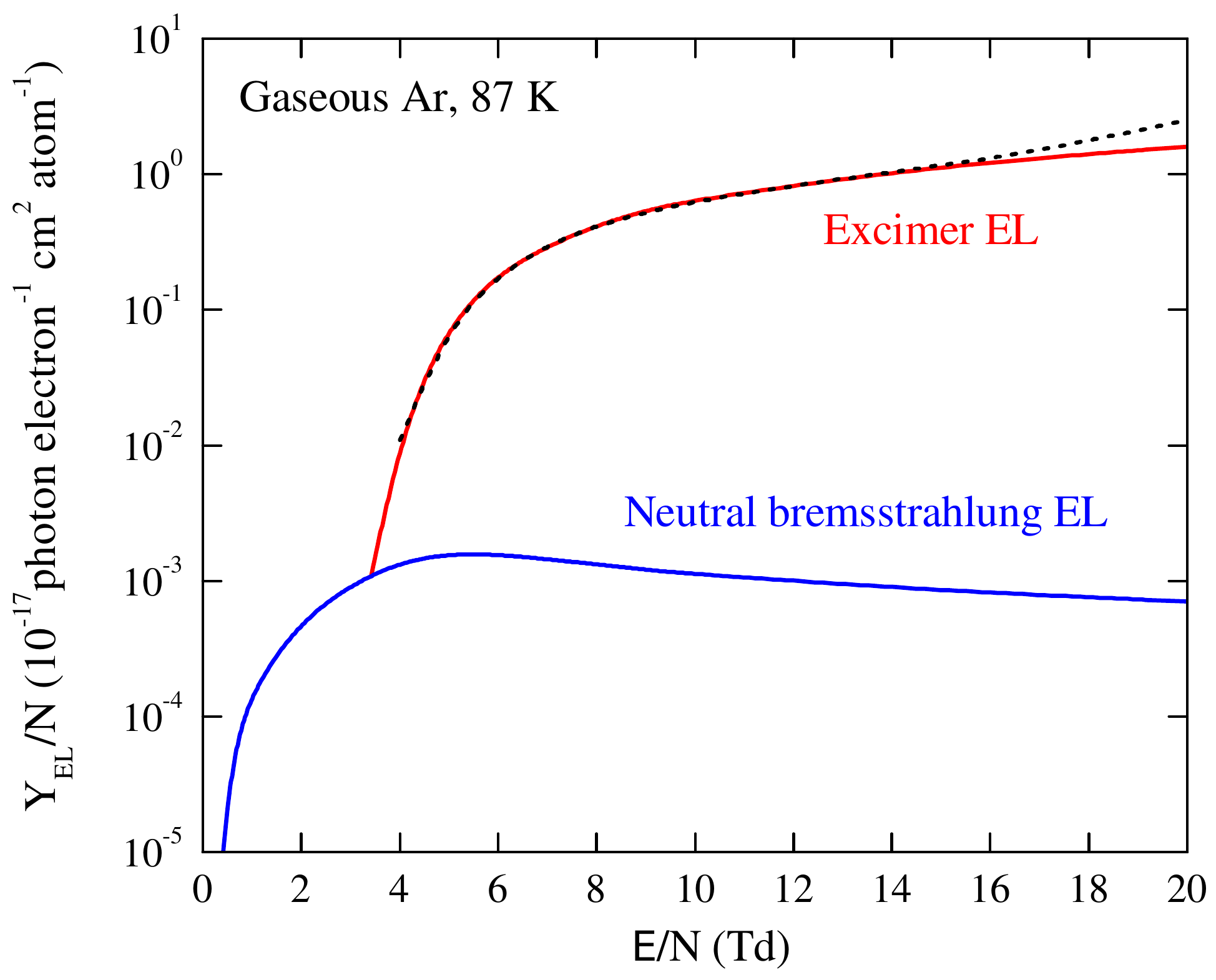}}\\
	\end{minipage}
	\hfill
	\begin{minipage}[h]{0.49\linewidth}
		\center{\includegraphics[width=1\linewidth]{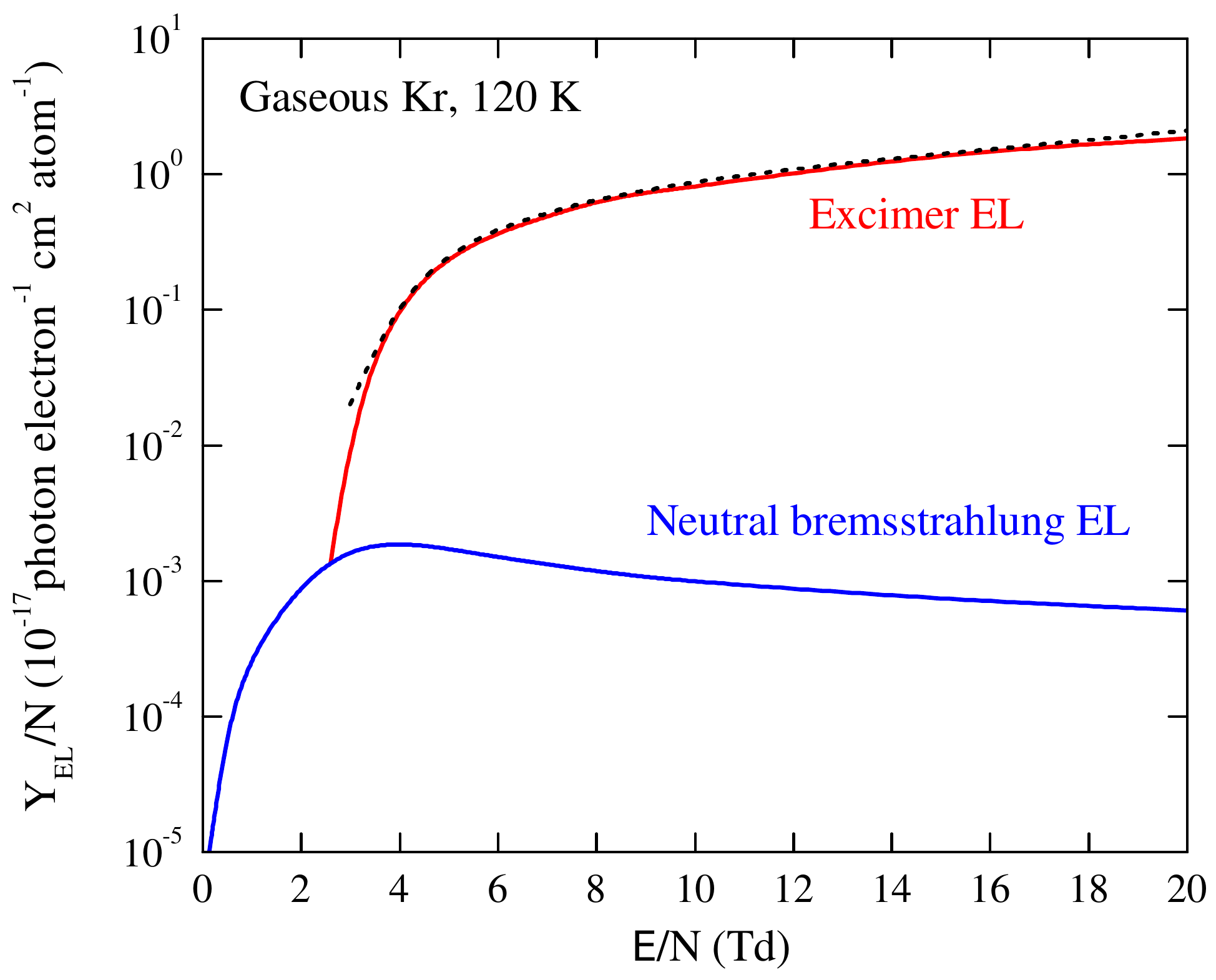}}\\
	\end{minipage}
	\vfill
	\begin{minipage}[h]{0.49\linewidth}
	\center{\includegraphics[width=1\linewidth]{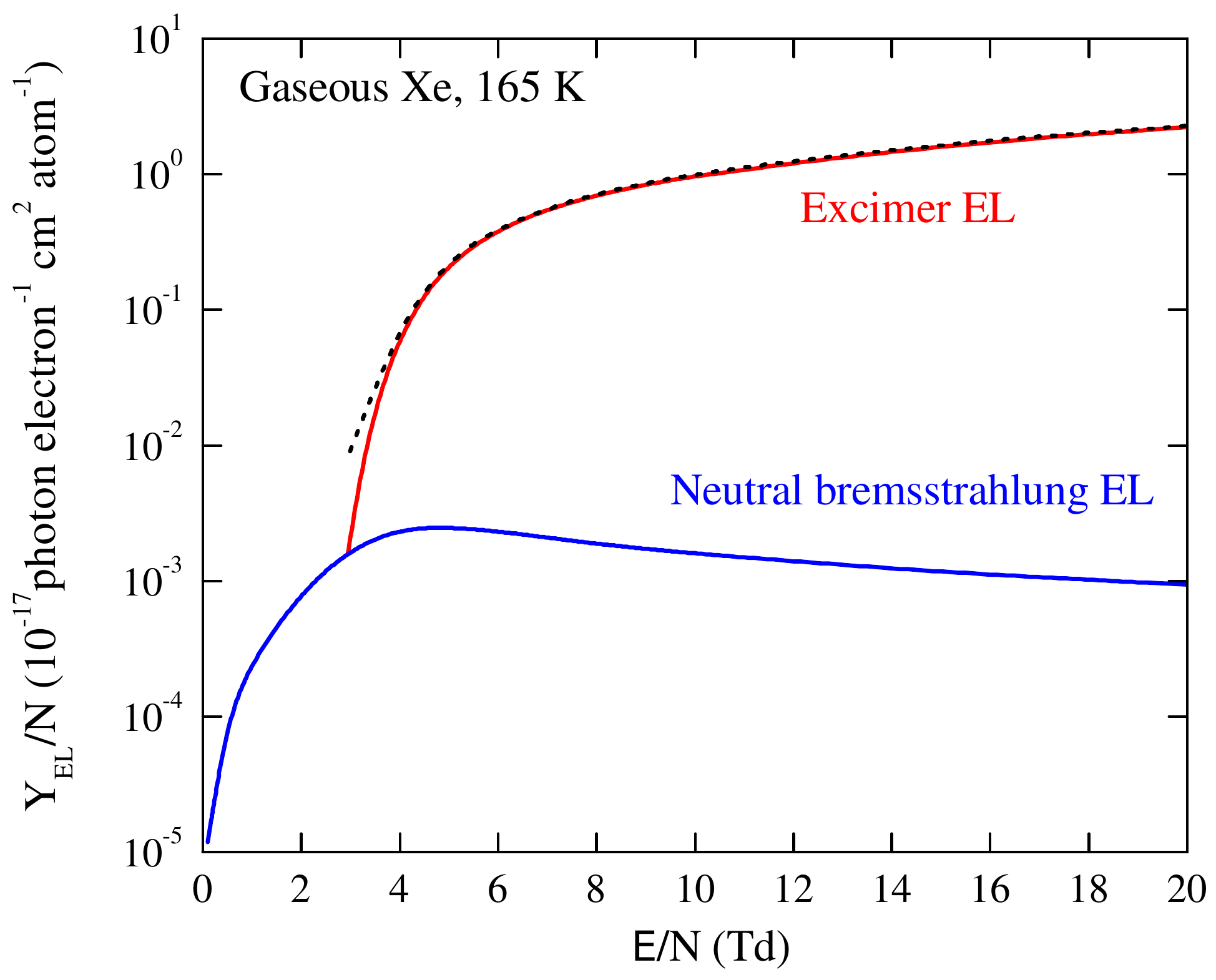}}\\
	\end{minipage}
	\vfill
	
	\caption{Reduced EL yield for NBrS EL at 0-1000 nm and that of excimer EL in noble gases as a function of the reduced electric field, calculated in this work using Boltzmann equation solver (solid lines).  For comparison, the EL yield of excimer EL, calculated using microscopic approach \cite{Oliveira11}, is shown (dashed lines).}
	\label{Fig10}
\end{figure*}

\begin{figure}[!h]
	\centering
	\includegraphics[width=1\linewidth]{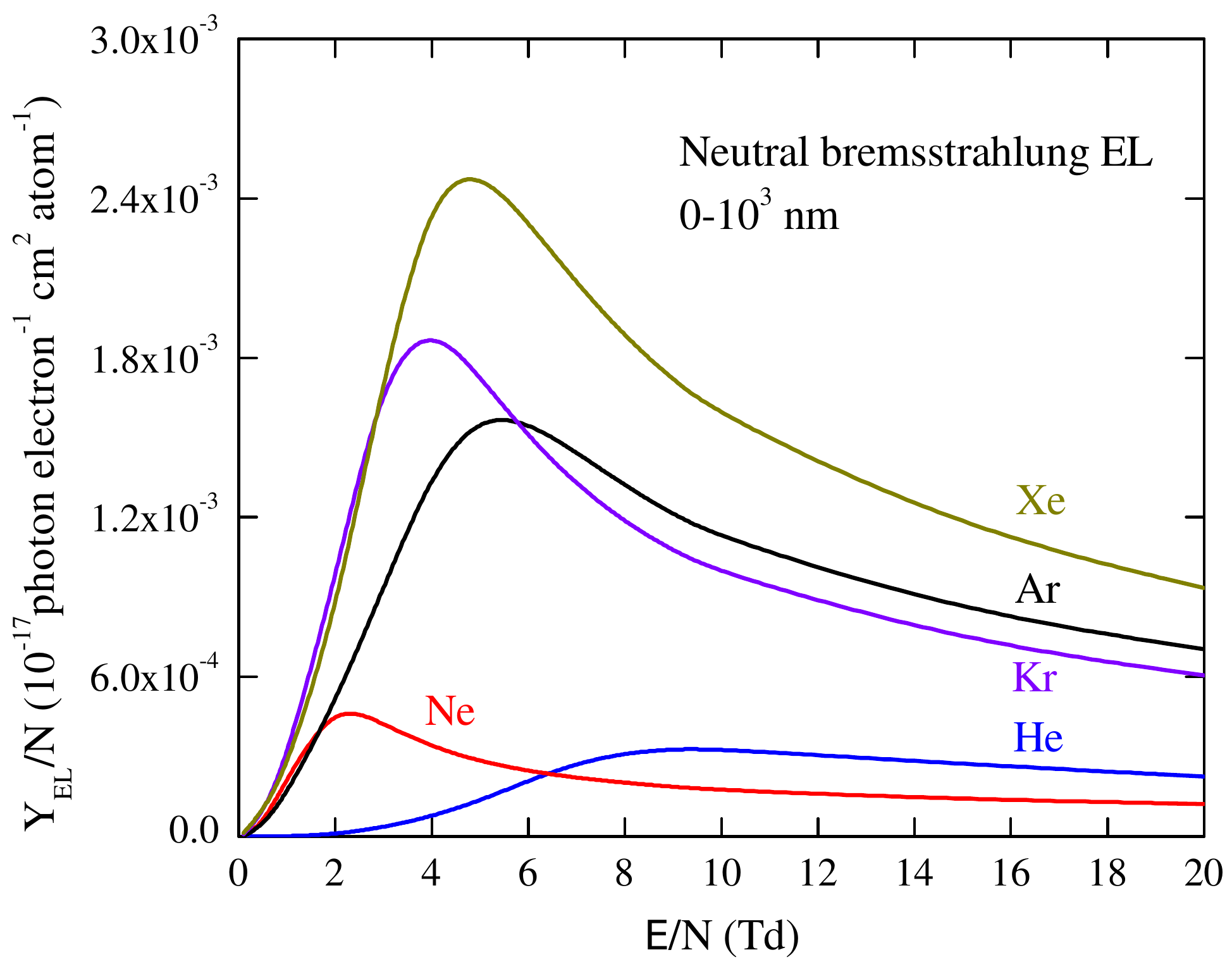}
	\caption{Reduced EL yield of NBrS EL in noble gases as a function of the reduced electric field, calculated using Eq.~\ref{Eq-NBrS-el-yield} integrated over the wavelength range of 0-1000 nm .}
	\label{Fig09}
\end{figure}

\begin{figure}[!h]
	\centering
	\includegraphics[width=1\linewidth]{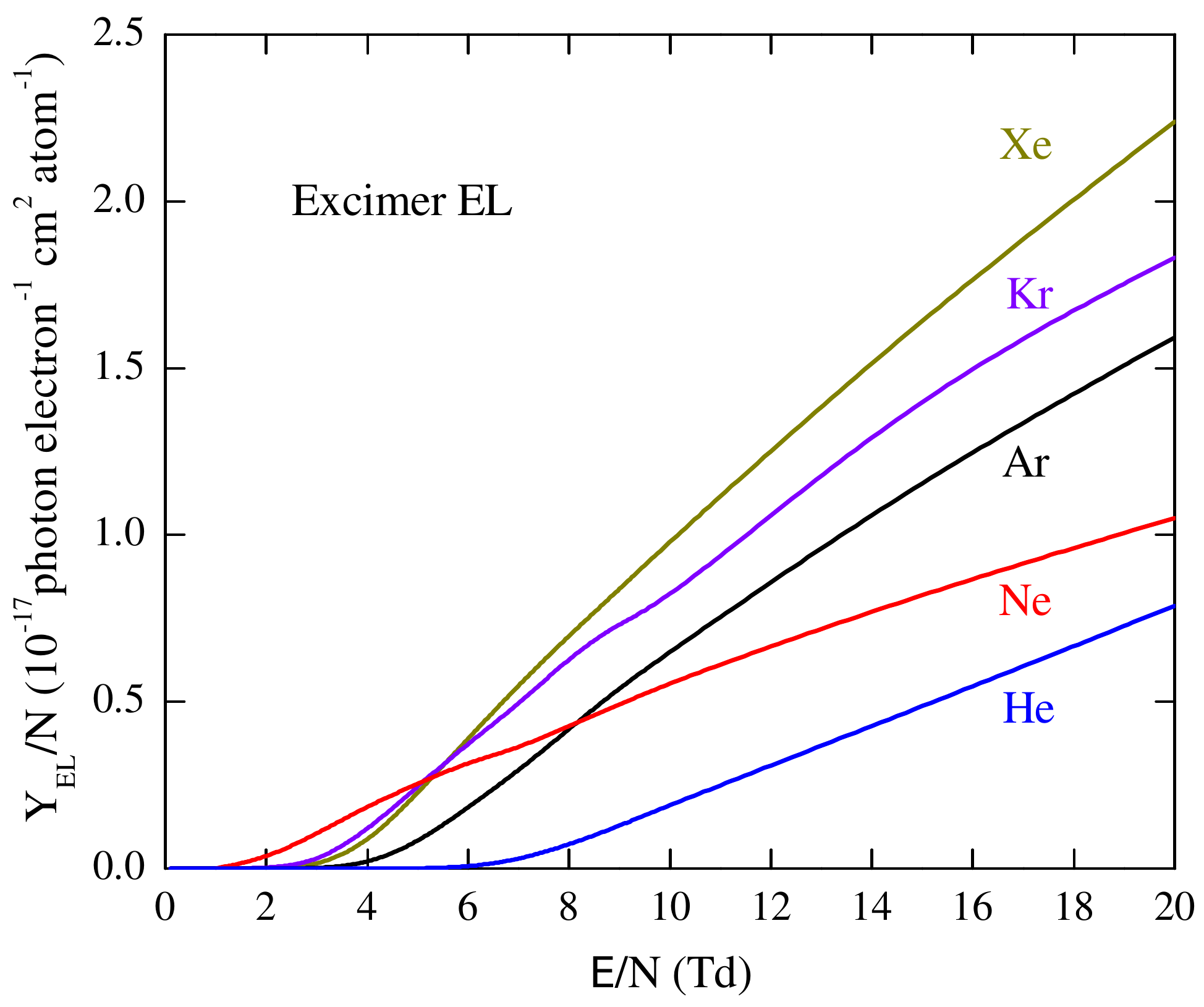}
	\caption{Reduced EL yield of excimer EL in noble gases as a function of the reduced electric field, calculated using Eq.~\ref{Eq-ord-el-yield}.}
	\label{Fig09a}
\end{figure}

The EL yields for both NBrS and excimer EL are presented in Figs.~\ref{Fig10}, \ref{Fig09} and \ref{Fig09a}, obtained by numerical integration of Eqs.~\ref{Eq-NBrS-el-yield} and \ref{Eq-ord-el-yield}. One can see from Fig.~\ref{Fig10} that the ratio between NBrS and excimer EL yields and their field dependence are generally the same for all noble gases. Details are described below.

Figs.~\ref{Fig10} and \ref{Fig09} shows the reduced EL yield of NBrS EL as a function of the reduced electric field, obtained by integration over the wavelength range of 0-1000 nm. The long-wavelength limit of 1000 nm is defined by that of SiPM sensitivity (see Fig.~\ref{FigArSpectra}). Thereby this yield corresponds to the maximum number of NBrS photons that can ever be detected by existing devices. 

One can see from Fig.~\ref{Fig10} and \ref{Fig09} that the NBrS EL yield first increases with the electric field, reaches a maximum at a certain field value ($\mathcal{E}/N_{max}$; see Table~\ref{tbl:table1},~item 6) and then slowly decreases. 
Such a behavior, at a slowly varying elastic cross section (shown in Fig.~\ref{Fig01}), reflects that of the $\upsilon_e/\upsilon_d$ ratio in Eq.~\ref{Eq-NBrS-el-yield}: see Fig.~\ref{Fig03}. In terms of  reduced EL yield, heavy noble gases are superior to light noble gases: the reduced yield of NBrS EL in Ar, Kr and Xe at maximum can exceed that of He and Ne by a factor of 5 (see Fig.~\ref{Fig09}). 

It should be remarked that the present theory of NBrS EL cannot explain the excess at higher fields of experimentally measured EL yields over theoretical prediction \cite{Buzulutskov18,Tanaka20}: see Fig.~\ref{Fig00Ar}. In Ar, the excess factor reaches 10 at a reduced field of 8 Td. As discussed in section 2, this excess might be explained by the additional EL mechanism, namely by NBrS on Feshbach and other negative ion resonances (see Eq.~\ref{NBrS-Res}), which is not accounted for by the present theory. Since Feshbach resonances exist in all noble gases (see Table~\ref{tbl:table1},~item 4), such an excess NBrS emission might exist in all noble gases as well. Further research is needed to clarify this issue.

The reduced EL yield of excimer EL is shown in  Figs.~\ref{Fig10} and \ref{Fig09a} for all noble gases. One can see that the yield in the first approximation increases linearly with the field. That is why excimer EL is called proportional. 

In addition, excimer EL has a well-defined threshold in the electric field, in contrast to NBrS EL the threshold of which tends to zero: compare Figs.~\ref{Fig09} and \ref{Fig09a}. The ``nominal'' excimer EL threshold can be defined as the intersection of extrapolation of the linear part of the curve with horizontal axis. The resulted thresholds are presented in Table~\ref{tbl:table1}~(item 7); these amount to 6.0, 1.5, 4.0, 3.0 and 3.5 Td for He, Ne, Ar, Kr and Xe respectively, in reasonable agreement with those obtained in microscopic approach \cite{Oliveira11}. It is interesting that due to interplay between the elastic and excitation cross sections, the minimum and maximum thresholds belong to Ne and He respectively. 

For a reduced electric field of 10 Td, the reduced yield for excimer EL increases in the serious He, Ne, Ar, Kr and Xe, the difference between He and Xe reaching a factor of 5 (see Fig.~\ref{Fig09a}). On the other hand, the reduced yield in Ne becomes equal to that of Ar at 8 Td and to that of Kr and Xe at 5 Td. This is because Ne has the lowest excimer EL threshold, of 1.5 Td.  

The ratio between the excimer and the NBrS EL yield can be deduced from Fig.~\ref{Fig10}: in each noble gas, it changes from about 1 at the nominal threshold of excimer EL to about 1000 at 10 Td. The latter might be reduced to about 100 if to take into account the NBrS enhancement at higher fields observed in experiment (see above).

The reliability of the results obtained in this work using the approach of Boltzmann equation solver can be checked by comparing to the calculations in the microscopic approach \cite{Oliveira11}. This comparison is done in Fig.~\ref{Fig10} for excimer EL yields in all noble gases, except He: there is a rather good compliance between the two approaches. The discrepancy at higher fields in Ne and Ar is due to the multiplication of electrons by impact ionization: in our approach we consider the EL yield per drifting electron, while in the microscopic approach the yield was taken for all electrons.

\section{Relevance to two-phase dark matter detectors}

In this section the relevance of the results obtained to two-phase detectors for dark matter search is discussed. We evaluate here the absolute EL yields, in terms of the number of photons produced by a drifting electron, and absolute electric fields and voltages needed to provide given reduced electric fields: these are presented in Table~\ref{tbl:table1}~(items 8-11). The estimations are given for a particular EL gap of a thickness of 1 cm, used in practice in two-phase Ar and Xe detectors \cite{Chepel13,Bernabei15,Aprile12}.

Using gas atomic densities of Table~\ref{tbl:table1}~(item 2), it can be deduced that the typical reduced electric fields in the EL gap,  in dark matter search experiments, were 7-10~Td in two-phase Xe detectors \cite{Chepel13,Bernabei15,Aprile12} and 4.6~Td in two-phase Ar detector \cite{Agnes15}. Higher reduced fields in two-phase Xe detectors are explained by the necessity to operate at higher extraction fields at the liquid-gas interface to provide the effective electron transmission through it \cite{Chepel13}.

The reduced electric field at which NBrS EL yield has a maximum predicted by the theory, $\mathcal{E}/N_{max}$, is about 1-1.5 Td higher than the threshold of excimer EL: see Table~\ref{tbl:table1}~(items 6 and 7). Therefore it is natural to compare the EL yields at just $\mathcal{E}/N_{max}$ where both EL mechanisms exist: see Table~\ref{tbl:table1}~(items 8 and 9). At such fields, the number of photons of NBrS EL in the two-phase mode is 1.2-1.6 per drifting electron per 1 cm in all noble gases, except He where it is about 8 due to considerably larger atomic density. 

On the other hand, the EL yield of excimer EL at these fields significantly exceeds that of NBrS EL predicted theoretically, by about a factor of 100. In fact, this large difference might be much less, of about a factor of 30, due to enhanced NBrS emission observed in experiment (see discussion in section 5). This difference might be further reduced, because in all noble gases (except Xe) it is necessary to use WLS to record excimer EL, the light losses in which can reach a factor of 10-20 \cite{Buzulutskov18}. 

Thus one may conclude that at moderate reduced electric fields, 1-1.5 Td above the excimer EL threshold, the direct optical readout of two-phase detectors  in the visible range, based on NBrS EL, can compete with the readout in the VUV with the use of WLS, based on excimer EL, in all noble gases, except Xe. 

Accordingly, NBrS EL is not practical for using in two-phase Xe detectors. On the other hand in other noble gases, NBrS EL can be used for the direct (without WLS) optical readout in the visible range, that may serve as a backup solution in two-phase detectors based on He, Ne, Ar and Kr, in case issues with WLS instability over time or non uniformity over large areas should become problematic.

Now let us estimate the electric fields and voltages for a 1 cm thick EL gap, operated in the two-phase mode in different noble gases: see Table~\ref{tbl:table1}~(items 10 and 11). The lighter the noble gas, the lower its boiling temperature ($T_b$),  and the higher its atomic density in the gas phase in the two-phase mode: see Table~\ref{tbl:table1}~(items 1 and 2). Thereby, the lighter the noble gas, the higher the absolute electric field needed to provide given reduced electric fields. In particular, item 10 of Table~\ref{tbl:table1} presents the electric field strength corresponding to the reduced electric field of 1 Td, while item 11 shows the voltage needed to provide the reduced electric field of 10 Td in a 1 cm thick EL gap. The latter amounts to  237, 34, 8.6, 6.2 and 5.8 kV for He, Ne, Ar, Kr and Xe respectively.

While for Ne, Ar, Kr and Xe this voltage is safe with respect to breakdowns, for He it exceeds the breakdown voltage in liquid He, which is about 100 kV~\cite{Gerhold94}. For He, a simple solution to the breakdown problem would be to reduce the EL gap thickness (and thus the gap voltage) by an order of magnitude, down to 1 mm: the number of photons produced in such a small gap for both NBrS and excimer EL would be still comparable to that for other noble gases: see Table~\ref{tbl:table1}~(items 8 and 9). Moreover, for such a small gap thickness the parallel-plate EL gap can be replaced by a more robust thick Gas Electron Multiplier (THGEM,~\cite{Breskin09}) having a similar thickness and operated in proportional EL mode. 

It should be remarked that for two-phase He and Ne detectors, the issue of electron emission from the liquid to the gas phase might be another problem, since the electrons in liquid He and Ne are trapped in bubbles, in contrast to other noble liquids. This resulted in that penetration of charges from the liquid to the gas phase in two-phase Ne is more complex and that the trapping time of the electrons at the liquid-gas interface is much larger, than expected \cite{Galea07}.  

These problems may force to give up of the two-phase mode for He- and Ne-based detectors and lead to the idea of a high-pressure single-phase cryogenic detector with EL gap, operated at temperatures slightly higher than $T_b$. Indeed, the gas atomic density at such low temperatures in Ne and especially in He can approach to the liquid atomic density in Ar, Kr, and Xe: see Table~\ref{tbl:table1}~(items 2 and 3), which makes it possible to abandon the liquid phase in He and Ne detectors at low temperatures and high ($\leq$10~atm) pressures. 

Finally, other prospects to use He and Ne media at low temperatures for dark matter and neutrino detectors should also be mentioned~\cite{Buzulutskov05,Ju07,Guo13,Liao21}.

\section{Conclusions}

In this work, the electroluminescence (EL) yields and spectra for both neutral bremsstrahlung (NBrS) and excimer  EL were calculated in He, Ne, Ar, Kr and Xe, following the theoretical approach successfully  applied to Ar in our previous work \cite{Buzulutskov18}. 

NBrS EL is predicted to have generally the same properties in all noble gases: 

- it has a rather flat emission spectrum extending  from the UV to the visible and NIR range, the vast majority of which is above 200 nm; 

- it dominates below the excimer EL threshold; 

- it can compete with excimer EL in terms of detected light intensity for moderate reduced electric fields, 1-1.5 Td above the excimer EL threshold;

- it is insignificant compared to excimer EL at higher reduced electric fields.

It was shown that in the two-phase mode, light noble gases (He and Ne) are as good as heavy noble gases (Ar, Kr and Xe) in terms of the number of photons (for both NBrS and excimer EL) emitted in a practical EL gap, of a thickness 1 mm for He and 1 cm for other noble gases. 

It was argued that NBrS EL might have practical applications in all noble gases, except Xe: it can be used for the direct (without WLS) optical readout in the visible range, that may serve as a backup solution for two-phase dark matter detectors, in case issues with WLS instabilities should become problematic.

\section*{Acknowledgments}

This work was supported by Russian Science Foundation (project no. 19-12-00008). It was done within the R\&D program of the DarkSide-20k experiment. 

\bibliographystyle{spphys_modified}       
\bibliography{mybibliography}   

\newcommand{\JINST}{J. Instrum.}\newcommand{\PhysLett}{Phys.
  Lett.}\newcommand{\IEEETransElectronDevices}{IEEE Trans. Electron
  Devices}\newcommand{\IEEETransNuclSci}{IEEE Trans. Nucl.
  Sci.}\newcommand{\AstropartPhys}{Astropart. Phys.}\newcommand{\NIM}{Nucl.
  Instrum. Methods}\newcommand{\NIMA}{Nucl. Instrum. Meth.
  A}\newcommand{\JApplPhys}{J. Appl. Phys.}\newcommand{\EPJWebConf}{EPJ Web
  Conf.}\newcommand{\EPJC}{Eur. Phys. J. C}\newcommand{\PhysRevLett}{Phys. Rev.
  Lett.}\newcommand{\PhysRevD}{Phys. Rev.
  D}\newcommand{\InstrumExpTech}{Instrum. Exp.
  Tech.}\newcommand{\InstrumMDPI}{Instruments}\newcommand{\JPAC}{J. Cosmol.
  Astropart. Phys.}
\begin{thebibliography}{10}
\providecommand{\url}[1]{{#1}}
\providecommand{\urlprefix}{URL }
\expandafter\ifx\csname urlstyle\endcsname\relax
  \providecommand{\doi}[1]{doi: \discretionary{}{}{}#1}\else
  \providecommand{\doi}{doi: \discretionary{}{}{}\begingroup
  \urlstyle{rm}\Url}\fi

\bibitem{Chepel13}
V.~Chepel{,} H.~Araujo, \JINST{} \textbf{8}, R04001 (2013).
\newblock \url{https://doi.org/10.1088/1748-0221/8/04/R04001}

\bibitem{Buzulutskov20}
A.~Buzulutskov, \InstrumMDPI{} \textbf{4}, 16 (2020).
\newblock \url{https://doi.org/10.3390/instruments4020016}

\bibitem{Buzulutskov18}
A.~Buzulutskov et~al., \AstropartPhys{} \textbf{103}, 29 (2018).
\newblock \url{https://doi.org/10.1016/j.astropartphys.2018.06.005}

\bibitem{Butikov70}
Y.~Butikov et~al., Sov. Phys. JETP \textbf{30}, 24 (1970)

\bibitem{Bondar20}
A.~Bondar et~al., \NIMA{} \textbf{958}, 162432 (2020).
\newblock \url{https://doi.org/10.1016/j.nima.2019.162432}

\bibitem{Monteiro08}
C.~Monteiro et~al., Phys. Lett. B \textbf{668}, 167 (2008).
\newblock \url{https://doi.org/10.1016/j.physletb.2008.08.030}

\bibitem{Buzulutskov11}
A.~Buzulutskov et~al., Europhys. Lett. \textbf{94}, 52001 (2011).
\newblock \url{https://doi.org/10.1209/0295-5075/94/52001}

\bibitem{Oliveira11}
C.~Oliveira et~al., \PhysLett{} B \textbf{703}, 217 (2011).
\newblock \url{https://doi.org/10.1016/j.physletb.2011.07.081}

\bibitem{Buzulutskov17}
A.~Buzulutskov, Europhys. Lett. \textbf{117}, 39002 (2017).
\newblock \url{https://doi.org/10.1209/0295-5075/117/39002}

\bibitem{Schwentner85}
N.~Schwentner, E.~Koch{,} J.~Jortner, \emph{Electronic Excitations in Condensed
  Rare Gases} (Springer, Berlin, 1985).
\newblock \url{https://doi.org/10.1007/BFb0111641}

\bibitem{Benson18}
C.~Benson et~al., \EPJC{} \textbf{78}, 329 (2018).
\newblock \url{https://doi.org/10.1140/epjc/s10052-018-5807-z}

\bibitem{Aprile06}
E.~Aprile, A.~Bolotnikov, A.~Bolozdynya{,} T.~Doke, \emph{Noble Gas Detectors}
  (Wiley, Weinheim, 2006).
\newblock \url{https://doi.org/10.1002/9783527610020}

\bibitem{Huffman65}
R.~Huffman et~al., Appl. Opt. \textbf{4}(12), 1581 (1965).
\newblock \url{https://doi.org/10.1364/AO.4.001581}

\bibitem{Morozov08}
A.~Morozov et~al., \JApplPhys{} \textbf{103}, 103301 (2008).
\newblock \url{https://doi.org/10.1063/1.2931000}

\bibitem{Lindblom88}
P.~Lindblom{,} O.~Solin, \NIMA{} \textbf{268}, 204 (1988).
\newblock \url{https://doi.org/10.1016/0168-9002(88)90607-9}

\bibitem{Fraga00}
M.~Fraga et~al., \IEEETransNuclSci{} \textbf{47}, 933 (2000).
\newblock \url{https://doi.org/10.1109/23.856721}

\bibitem{Aalseth21}
C.~Aalseth{ et al. [DarkSide-20k collaboration]}, \EPJC{} \textbf{81}, 153
  (2021).
\newblock \url{https://doi.org/10.1140/epjc/s10052-020-08801-2}

\bibitem{Oliveira13}
C.~Oliveira et~al., \NIMA{} \textbf{722}, 1 (2013).
\newblock \url{https://doi.org/10.1016/j.nima.2013.04.061}

\bibitem{Tanaka20}
M.~Tanaka et~al., J. Phys.: Conf. Series \textbf{1468}, 012052 (2020).
\newblock \url{https://doi.org/10.1088/1742-6596/1468/1/012052}

\bibitem{Kimura20}
M.~Kimura et~al., {\JINST{}} \textbf{15}(08), C08012 (2020).
\newblock \url{https://doi.org/10.1088/1748-0221/15/08/C08012}

\bibitem{Takeda20}
T.~Takeda et~al., J. Phys.: Conf. Series \textbf{1468}, 012053 (2020).
\newblock \url{https://doi.org/10.1088/1742-6596/1468/1/012053}

\bibitem{Takeda20a}
T.~Takeda et~al., \JINST{} \textbf{15}(03), C03007 (2020).
\newblock \url{https://doi.org/10.1088/1748-0221/15/03/c03007}

\bibitem{Aoyama21}
K.~Aoyama et~al., eprint arXiv: 2107.02330  (2021).
\newblock \url{https://arxiv.org/abs/2107.02330}

\bibitem{Schulz73}
G.~Schulz, Rev. Mod. Phys. \textbf{45}, 378 (1973).
\newblock \url{https://doi.org/10.1103/RevModPhys.45.378}

\bibitem{Dyachkov74}
L.~D'yachkov et~al., Sov. Phys. JETP \textbf{38}, 697 (1974).
\newblock \url{https://www.osti.gov/biblio/4400199}

\bibitem{Monteiro21}
C.~Monteiro, Novel approach to Xenon optical TPCs: the presence of Neutral
  Bremsstrahlung, TIPP Conf., May 24-28, 2021, Canada, Vancouver
  \url{https://indi.to/m5qnK}

\bibitem{Park00}
J.~Park et~al., Phys. Plasmas \textbf{7}(8), 3141 (2000).
\newblock \url{https://doi.org/10.1063/1.874220}

\bibitem{Firsov61}
O.~Firsov{,} M.~Chibisov, Sov. Phys. JETP \textbf{12}, 1235 (1961).
\newblock \url{https://www.osti.gov/biblio/4094108}

\bibitem{Kasyanov65}
V.~Kas'yanov{,} A.~Starostin, Sov. Phys. JETP \textbf{21}, 193 (1965)

\bibitem{Dalgarno66}
A.~Dalgarno{,} N.~Lane, Astrophys. J. \textbf{145}, 623 (1966).
\newblock \url {http://dx.doi.org/10.1086/148801}

\bibitem{Biberman67}
L.~Biberman{,} G.~Norman, Sov. Phys. Uspekhi \textbf{10}, 52 (1967).
\newblock \url {https://doi.org/10.1070/pu1967v010n01abeh003199}

\bibitem{Bolsig1}
G.~Hagelaar{,} L.~Pitchford, Plasm. Sour. Sci. Tech. \textbf{14}, 722 (2005).
\newblock \url {https://doi.org/10.1088/0963-0252/14/4/011}

\bibitem{Bolsig2}
https://fr.lxcat.net/solvers/bolsigplus/

\bibitem{DBBiagi}
www.lxcat.net/biagi

\bibitem{Biagi99}
S.~Biagi, \NIMA{} \textbf{421}(1), 234 (1999).
\newblock \url{https://doi.org/10.1016/S0168-9002(98)01233-9}

\bibitem{DBBSR}
www.lxcat.net/bsr

\bibitem{Peisert84}
A.~Peisert{,} F.~Sauli, \emph{{Drift and diffusion of electrons in gases: a
  compilation}}.
\newblock CERN Yellow Reports: Monographs (CERN, Geneva, 1984).
\newblock \url{https://doi.org/10.5170/CERN-1984-008}

\bibitem{Kasyanov78}
V.~Kas'yanov{,} A.~Starostin, Sov. J. Plasma Phys. \textbf{4}, 67 (1978).
\newblock \url{https://www.osti.gov/biblio/6666613}

\bibitem{Fastovsky71}
V.~Fastovsky, A.~Rovinsky{,} Y.~Petrovsky, \emph{Inert Gases (in Russian)}
  (Atomizdat, Moscow, 1971)

\bibitem{Theeuwes70}
F.~Theeuwes{,} R.~Bearman, J. Chem. Thermodynamics \textbf{2}, 507 (1970).
\newblock \url{https://doi.org/10.1016/0021-9614(70)90100-X}

\bibitem{Bernabei15}
R.~Bernabei et~al., Int. J. Mod. Phys. A \textbf{30}, 1530053 (2015).
\newblock \url{https://doi.org/10.1142/S0217751X15300537}

\bibitem{Aprile12}
E.~Aprile et~al., \AstropartPhys{} \textbf{35}, 573 (2012).
\newblock \url {https://doi.org/10.1016/j.astropartphys.2012.01.003}

\bibitem{Agnes15}
P.~Agnes et~al., \PhysLett{} B \textbf{743}, 456 (2015).
\newblock \url{https://doi.org/10.1016/j.physletb.2015.03.012}

\bibitem{Gerhold94}
J.~Gerhold et~al., Cryogenics \textbf{34}, 579 (1994).
\newblock \url{https://doi.org/10.1016/0011-2275(94)90183-X}

\bibitem{Breskin09}
A.~Breskin et~al., \NIMA{} \textbf{598}, 107 (2009).
\newblock \url{https://doi.org/10.1016/j.nima.2008.08.062}

\bibitem{Galea07}
R.~Galea et~al., JINST \textbf{2}, P04007 (2007).
\newblock \url{https://doi.org/10.1088/1748-0221/2/04/P04007}

\bibitem{Buzulutskov05}
A.~Buzulutskov et~al., \NIMA{} \textbf{548} (2005).
\newblock \url{https://doi.org/10.1016/j.nima.2005.04.066}

\bibitem{Ju07}
Y.~Ju et~al., Cryogenics \textbf{47}, 81 (2007).
\newblock \url{https://doi.org/10.1016/j.cryogenics.2006.08.008}

\bibitem{Guo13}
W.~Guo{,} D.~Mckinsey, Phys. Rev. D \textbf{87}, 115001 (2013).
\newblock \url{https://doi.org/10.1103/PhysRevD.87.115001}

\bibitem{Liao21}
J.~Liao et~al., eprint arXiv: 2103.02161  (2021).
\newblock \url{https://arxiv.org/abs/2103.02161}

\end{thebibliography}

%
%

\end{document}